\documentclass{aa}  

\usepackage{graphicx}
\usepackage[varg]{txfonts}
\usepackage{hyperref}
\usepackage{xcolor, subcaption, textcomp}
\newcommand\T{\rule{0pt}{2.6ex}}       
\newcommand\B{\rule[-1.2ex]{0pt}{0pt}} 
\begin{document}

   \title{Successive deuteration in low-mass star-forming regions:\\the case of D$_{2}$-methanol (CHD$_{2}$OH) in IRAS~16293-2422}
	 \titlerunning{Successive deuteration in low-mass star-forming regions}

   \author{Maria N. Drozdovskaya\inst{1}
          \and
					Laurent H. Coudert\inst{2}
					\and
					Laurent Margul\`{e}s\inst{3}
					\and
					Audrey Coutens\inst{4}
					\and\\
          Jes K. J\o{}rgensen\inst{5}
					\and
					S\'{e}bastien Manigand\inst{6}
          }

   \institute{Center for Space and Habitability, Universit\"{a}t Bern, Gesellschaftsstrasse 6, CH-3012 Bern, Switzerland\\
              \email{maria.drozdovskaya@unibe.ch; maria.drozdovskaya.space@gmail.com}
         \and
				     Institut des Sciences Mol\'{e}culaires d'Orsay (ISMO), CNRS, Universit\'{e} Paris-Saclay, F-91405 Orsay, France\\
              \email{laurent.coudert@universite-paris-saclay.fr}
				 \and
				     Laboratoire de Physique des Lasers, Atomes, et Mol\'ecules, UMR CNRS 8523, Universit\'e de Lille, F-59655 Villeneuve d'Ascq C\'edex, France\\
						  \email{laurent.margules@univ-lille.fr}
				 \and
				     Institut de Recherche en Astrophysique et Plan\'{e}tologie, Universit\'{e} de Toulouse, UPS-OMP, CNRS, CNES, 9 av. du Colonel Roche, 31028 Toulouse Cedex 4, France\\
						  \email{audrey.coutens@irap.omp.eu}
				 \and
             Niels Bohr Institute, University of Copenhagen, \O{}ster Voldgade 5-7, DK-1350 Copenhagen K., Denmark\\
						  \email{jeskj@nbi.ku.dk}
				 \and
				     Laboratoire d'Etudes Spatiales et d'Instrumentation en Astrophysique (LESIA), Observatoire de Paris, Universit{\'e} PSL, CNRS, Sorbonne Universit{\'e}, Universit{\'e} de Paris, 5 place Jules Janssen, 92195 Meudon, France
\\
						  \email{sebastien.manigand@obspm.fr}
             }

   \date{Received 2021; accepted 18 January 2022}

 
  \abstract
   {Di-deuterated molecules are observed in the earliest stages of star formation at abundances of a few \% relative to their non-deuterated isotopologs, which is unexpected considering the scarcity of deuterium in the interstellar medium. With sensitive observations leading to the detection of a steadily increasing number of di-deuterated species, it is becoming possible to explore successive deuteration chains.}
   {Accurate quantification of the column density of di-deuterated methanol is a key missing puzzle piece in the otherwise thoroughly constrained family of D-bearing methanol in the deeply embedded low-mass protostellar system and astrochemical template source IRAS~16293-2422. A spectroscopic dataset for astrophysical purposes is built for CHD$_{2}$OH and made publicly available to facilitate accurate characterization of this species in astrochemical surveys.}
   {The newly computed line list and partition function are used to search for CHD$_{2}$OH towards IRAS~16293-2422~A and~B in data from the Atacama Large Millimeter/submillimeter Array (ALMA) Protostellar Interferometric Line Survey (PILS). Only non-blended, optically thin lines of CHD$_{2}$OH are used for the synthetic spectral fitting.}
   {The constructed spectroscopic database contains line frequencies and strengths for $7~417$ transitions in the $0$ to $500$~GHz frequency range. ALMA-PILS observations in the $329-363$~GHz range are used to identify $105$ unique, non-blended, optically thin line frequencies of CHD$_{2}$OH for synthetic spectral fitting. The derived excitation temperatures and column densities yield high D/H ratios of CHD$_{2}$OH in IRAS~16293-2422~A and~B of $7.5\pm1.1\%$ and $7.7\pm1.2\%$, respectively.}
   {Deuteration in IRAS~16293-2422 is not higher than in other low-mass star-forming regions (L483, SVS13-A, NGC~1333-IRAS2A, -IRAS4A, and -IRAS4B). Di-deuterated molecules consistently have higher D/H ratios than their mono-deuterated counterparts in all low-mass protostars, which may be a natural consequence of H-D substitution reactions as seen in laboratory experiments. The Solar System's natal cloud, as traced by comet 67P/Churyumov–Gerasimenko, may have had a lower initial abundance of D, been warmer than the cloud of IRAS~16293-2422, or been partially reprocessed. In combination with accurate spectroscopy, a careful spectral analysis, and a consideration of the underlying assumptions, successive deuteration is a robust window on the physicochemical provenance of star-forming systems.}


   \maketitle
%

\section{Introduction}
\label{introduction}
Di-deuterated molecules are supposed to be scarce, since hydrogen is much more abundant than deuterium in the interstellar medium (ISM). The elemental abundance of deuterium relative to hydrogen in the local (within $\sim1-2$~kpc of the Sun) ISM has been derived to be at least $(2.0\pm0.1)\times10^{-5}$, including observed variations caused by the depletion of gas-phase deuterium onto dust grains \citep{Prodanovic2010}. However, the inventory of species containing two deuterium atoms is steadily increasing: D$_{2}$O \citep{Butner2007}, D$_{2}$CO \citep{Turner1990}, CHD$_{2}$OH \citep{Parise2002}, CHD$_{2}$OCHO \citep{Manigand2019}, CH$_{3}$OCHD$_{2}$ \citep{Richard2021}, CHD$_{2}$CCH and CH$_{2}$DCCD \citep{Agundez2021b}, NHD$_{2}$ \citep{Roueff2000}, ND$_{2}$ radical \citep{Bacmann2020}, CHD$_{2}$CN \citep{Calcutt2018a}, D$_{2}$S \citep{Vastel2003}, D$_{2}$CS \citep{Marcelino2005}, c-C$_{3}$D$_{2}$ \citep{Spezzano2013}, D$_{2}$H$^{+}$ \citep{Vastel2004}. This inventory of D$_{2}$-bearing molecules includes simple and complex organic species. Two tri-deuterated species have also been detected: ND$_{3}$ \citep{vanderTak2002} and CD$_{3}$OH \citep{Parise2004}. Moreover, the abundances of multiply deuterated isotopologs are typically high, at a level of a few \% relative to the normal (i.e., non-deuterated) species (e.g., \citealt{CaselliCeccarelli2012, CeccarelliPPVI}).

The low ($\sim10$~K) temperatures of dense ($>10^{4}$~cm$^{-3}$) cores are thought to be the sole environments capable of boosting chemical pathways towards D-bearing molecules via gas-phase and solid-state reactions \citep{Watson1974, Tielens1983, DalgarnoLepp1984, RodgersCharnley2002, Millar2003, Caselli2019b}. Di-deuterated molecules have been detected in starless and prestellar cores (e.g., \citealt{Roueff2000, Bacmann2003, Bergman2011a}), as well as in hot corinos of embedded low-mass protostars (e.g., \citealt{Ceccarelli1998c, Loinard2000, Ceccarelli2001}) and hot cores of high-mass star-forming regions (e.g., \citealt{Turner1990, Zahorecz2017, Zahorecz2021}). It is thought that D-bearing molecules observed in protostellar regions have formed in the preceding cold prestellar phases and preserved as ices until the later protostellar stage, when they undergo thermal desorption and become observable as gases (e.g., \citealt{Cazaux2011b, Aikawa2012b, Taquet2012b, Taquet2013a, Taquet2014}). Di-deuterated molecules have recently been detected in comets of our Solar System as well: D$_{2}$O \citep{Altwegg2017a} and di-deuterated methanol (although the exact isotopolog could not be identified; \citealt{Drozdovskaya2021}).

IRAS~16293-2422 is a frequently studied, well-characterized nearby ($141$~pc; \citealt{Dzib2018}) low-mass protostellar system. An extensive description of the source is available in \citet{Jorgensen2016}. It is thought to be a triple system of a single source B and a tight binary of A1 and A2. B is separated from the binary A by $5.3\arcsec$ ($\sim747$~au; \citealt{vanderWiel2019}). A1 and A2 have a separation of just $0.38\arcsec$ ($\sim54$~au; \citealt{Maureira2020a}). The proximity of IRAS~16293-2422 enables spatially resolved observations at high sensitivity to be carried out, making it an optimal target for searches for minor chemical constituents. Early-on, multiply deuterated species have been detected towards this source with IRAM~30~m observations: D$_{2}$CO \citep{Ceccarelli1998c}, CHD$_{2}$OH \citep{Parise2002}, CD$_{3}$OH \citep{Parise2004}. Now, with ALMA their precise spatial distribution within this triple system can be examined, thereby allowing accurate abundances to be determined in every individual physical component (circumbinary envelope vs. individual protostellar envelopes vs. central individual hot corinos) of the system. The ALMA-PILS survey of this source also revealed new (previously undetected in the ISM) di-deuterated molecules: CHD$_{2}$CN \citep{Calcutt2018a}, CHD$_{2}$OCHO \citep{Manigand2019}, and CH$_{3}$OCHD$_{2}$ \citep{Richard2021}.

Up to now, only single-dish estimates of the column density of CHD$_{2}$OH in IRAS~16293-2422 have been available in the literature \citep{Parise2002}. Moreover, all interstellar detections of this molecule thus far have been based on spectroscopic data of low quality. In this paper, new accurate spectroscopic data (Sect.~\ref{spectroscopy}) are used to firmly quantify the availability of CHD$_{2}$OH in IRAS~16293-2422 A and B (hereafter, I16293A and I16293B, respectively) on the basis of ALMA-PILS observations (Sect.~\ref{observations}). The catalog entry is also being made publicly available to enable accurate determination of di-deuterated methanol abundances in other sources. This work makes use of newly developed MCMC spectral fitting routines for the derivation of the column density and excitation temperature of a species (Sect.~\ref{spectral_analysis}). With the accurate column density of CHD$_{2}$OH in IRAS~16293-2422 on hand (Sect.~\ref{results}), the deuteration chains of all multiply deuterated molecules are investigated. A comparison is performed to other low-mass star-forming systems, as well as laboratory experiments and cometary measurements (Sect.~\ref{discussion}). The conclusions of this work are summarized in Sect.~\ref{conclusions}.


\section{Methods}
\label{methods}
\subsection{Observations}
\label{observations}

IRAS~16293-2422 was the target of the large unbiased Protostellar Interferometric Line Survey (PILS\footnote{\url{http://youngstars.nbi.dk/PILS/}}; project-id: 2013.1.00278.S, PI: Jes K. J\o{}rgensen) carried out with ALMA in Cycle~$2$ as presented in an extensive series of papers. These data cover the $329-363$~GHz frequency range (ALMA Band~$7$) at a spatial resolution of $\sim0.5\arcsec$ ($70$~au; the applied uniform circular restoring beam). This allows for sources A and B to be studied individually; however, this is not enough to spatially separate A1 from A2. The observations are a combination of the 12m array and the 7m Atacama Compact Array (ACA), hence sampling extended structures up to $\sim13\arcsec$ ($\sim2~000$~au; largest recoverable size). The data were obtained at a spectral resolution of $0.2$~km~s$^{-1}$ ($0.244$~MHz) and reached a sensitivity of $\sigma=8-10$~mJy~beam$^{-1}$~channel$^{-1}$ (or $4-5$~mJy~beam$^{-1}$~km~s$^{-1}$). For the typical line widths of $\sim1.0$ and $\sim2.2$~km~s$^{-1}$ towards favorable positions near sources B and A (with narrowest lines, least dust absorption, but yet still probing hot dense inner regions), respectively, this allowed the lines to be spectrally resolved by $\sim5-11$ spectral bins. The observations did not yet reach the line confusion limit of this source at these frequencies. The full details about the data set and its reduction are available in \citet{Jorgensen2016}.

\subsection{Spectroscopy of CHD$_{2}$OH}
\label{spectroscopy}

\begin{table}
\caption{Partition function $Q_{\text{rot-tors}}$ of CHD$_{2}$OH.}
\label{tbl:Qvalues}
\centering
\begin{tabular}{rr}
\hline
\hline
$T$ (K) & $Q_{\text{rot-tors}}$\T\B\\
\hline
2.725 & 4.6\T\\
5.000 & 12.4\\
9.375 & 39.8\\
18.75 & 145.3\\
37.50 & 490.5\\
75.00 & 1~563.4\\
150.0 & 5~255.9\\
225.0 & 11~268.4\\
300.0 & 19~423.4\B\\
\hline
\end{tabular}
\tablefoot{The partition function is given for each temperature $T$ in Kelvin.}
\end{table}

Di-deuterated methanol CHD$_{2}$OH is a non-rigid species displaying a large amplitude internal rotation of its methyl group. For this reason, it displays a complicated microwave spectrum consisting of many lines with no easily recognizable features. Perturbations take place even for low values of the rotational quantum number $K$, which prevented the first investigators from accurately predicting transition frequencies. In this investigation, torsional states are identified using either the labeling scheme of \citet{SuQuade1989b} or the torsional quantum number $v_{t}$. The one-to-one correspondence between both schemes is given in table~1 of \citet{Ndao2016}.

The first measurements of the microwave spectrum were reported by \citet{SuQuade1989a}. Thanks to a then new theoretical approach \citep{Quade1967} aimed at accounting for the internal rotation of an asymmetrical methyl group and with the help of the double resonance technique, parallel $a$-type transitions within the three lowest-lying torsional states, $e_{0}$, $o_{1}$, and $e_{1}$, could be assigned up to $J=2$ and $K=1$, as well as $\Delta K=\pm1$ perpendicular transitions with $J\leq9$ and $K\leq2$ involving the same torsional states. Thanks to a modified version of the theoretical approach \citep{LiuQuade1991a}, new parallel and perpendicular transitions could later be assigned and fitted \citep{LiuQuade1991b, FuSu1991a} up to $J=7$, $K=2$, and $o_{3}$. The number of assigned transitions further increased thanks to \citet{FuSu1991b} who were able to assign additional transitions within the first three torsional states, as well as perpendicular transitions within $e_{2}$ and $o_{2}$. Parallel transitions within the higher lying $o_{3}$ and $e_{3}$ were reported by \citet{Quade1998} and their assignments were confirmed with combination differences. An improved theoretical approach proposed by \citet{Ndao2016} led to more accurate rotation-torsion energy levels and allowed these authors to assign $3~271$ parallel and perpendicular transitions up to $J=40$, with $3\leq K\leq9$, within the three first torsional states. All transitions could be fitted using a phenomenological $J(J+1)$ expansion, which substantiated the assignments. It was also found that such an expansion fails for transitions with $0\leq K<3$ because the corresponding low-lying rotation-torsion energy levels are close together and perturbed. The theoretical approach \citep{Ndao2016} was used to predict the subband center of many torsional subbands characterized by a value of the torsional quantum number $v_{t}$ as high as $11$, which could be observed in the far infrared (FIR). The analysis of their rotational structure also substantiated the microwave transitions assignments. Independently and at the same time, I. Mukhopadhyay reported assignments for three sets of transitions. The first set \citep{Mukhopadhyay2016a} concerns parallel microwave and submillimeter wave transitions with $0\leq K\leq7$ assigned for the first three torsional states up to $J=8$. This first set also concerns $4$ perpendicular subbands involving the same torsional states, with $3\leq K\leq6$, assigned in submillimeter wave and FIR spectra. These perpendicular transitions were also reported by \citet{Ndao2016}. The second \citep{Mukhopadhyay2016b} and third \citep{Mukhopadhyay2016c} sets consist of perpendicular transitions assigned in submillimeter wave and FIR spectra: within the ground $e_{0}$ torsional state and up to $K=14$ for the first set and between the first three torsional states and up to $K=13$ for the third set. The torsional subbands with $K\leq9$ were also reported by \citet{Ndao2016}. In a subsequent paper by \citet{Mukhopadhyay2021}, additional FIR perpendicular subbands were reported for $3\leq K\leq7$ and up to $e_{5}$. Parallel and perpendicular microwave transitions were also reported for $K=3$ and $4$. Those involving the second and third torsional states were already reported in  \citet{Ndao2016}. The assignment of the perturbed transitions with $0\leq K<3$ within the three first torsional state was successfully carried out by \citet{Coudert2021} thanks to an improved version of the theoretical approach of \citet{Ndao2016}. For the same torsional torsional states, new parallel and perpendicular transitions were assigned up to $K=12$ and $J=40$. The first global line position analysis, restricted to transitions with $J\leq26$, was also performed.

The results of the global line position analysis of \citet{Coudert2021} were used to build a spectroscopic database for CHD$_{2}$OH. All rotation-torsion lines within and between the first three torsional states, $0\leq v_{t}\leq2$, were computed up to $J=26$. Line strengths were evaluated taking a dipole moment function with two non-vanishing constant components along the $x$ and $z$ axes of the molecule-fixed axis system such that the $xz$ plane remains parallel to the COH group. The values are $\mu_{x}=-1.37$ and $\mu_{z}=0.8956$~D as reported by \citet{Amano1981}. The partition function $Q_{\text{rot-tors}}$, which contains the rotational and torsional contributions, was calculated by taking a degeneracy factor equal to $(2J+1)$ and a zero energy for the lowest-lying rotation-torsion energy level, which is the $J=0$, $e_{0}$ level. It should be noted that the vibrational contribution to the partition function is negligible, because the vibrational modes are above $1000$~cm$^{-1}$. The values computed for temperatures ranging from $2.75$ to $300$~K are listed in Table~\ref{tbl:Qvalues}. Rotation-torsion lines with a frequency below $500$~GHz were selected and, as in the Jet Propulsion Laboratory (JPL) database catalog line files \citep{Pickett1998}, an intensity cutoff depending on the line frequency was taken. Its value in $\text{nm}^{2}\cdot\text{MHz}$ units at $300$~K is
\begin{equation}
10^{\mbox{{\small LOGSTR0}}} + (F/300~000)^{2} \times 10^{\mbox{{\small LOGSTR1}}},
\label{CUTOFF}
\end{equation}
where $F$ is the frequency in MHz, and {\small LOGSTR0} and {\small LOGSTR1} are two dimensionless constants both set to $-9$. The line list is published as an electronic Table at the Centre de Donn\'{e}es astronomiques de Strasbourg (CDS) with association to this A\&A article. It is formatted in the same way as the catalog line files of the JPL database \citep{Pickett1998} and displays $16$ columns. Columns $1$ to $3$ contain, respectively, the line frequency (FREQ) in MHz, the error (ERR) in MHz, and the base-$10$ logarithm of the line intensity (LGINT) in $\text{nm}^{2}\cdot\text{MHz}$ units at $300$~K. Columns $4$ to $6$ give the degrees of freedom of the rotational partition function (DR), the lower state energy (ELO) in cm$^{-1}$, and the upper state degeneracy (GUP), respectively. Columns $7$ and $8$ contain the species tag (TAG) and format number (QNFMT), respectively. Finally, columns $9$ to $12$ ($13$ to $16$) give the assignment of the upper (lower) level in terms of the rotational quantum numbers $J$, $K$, $p$ introduced by \citet{Coudert2021} and the torsional quantum number $v_{t}$. A minimum value of $10$~kHz was selected for the calculated error (ERR). For observed non-blended microwave lines, the line frequency (FREQ) and the error (ERR) were replaced by their experimental values. This is then indicated by a negative species tag.

A comparison between the present spectroscopic database and a previously available line list is shown in Table~\ref{tbl:comp_dtb}. For the $184$ parallel transitions given in table~$2$ of \citet{Mukhopadhyay2016a}, Table~\ref{tbl:comp_dtb} of the present paper lists the $32$ transitions with a frequency discrepancy larger than $0.5$~MHz. These discrepancies are due to assignment errors and can be as large as $788$~MHz. For instance, the parallel line at $333~159$~MHz initially assigned as a $K=6$, $e_{0}$ line turned out to be a $K=7$, $o_{1}$ line. The assignments of the closely lying lines at $125~031$ and $125~018$~MHz need to be exchanged.

\begin{table*}
\caption{Comparison between the database built in Sect.~\ref{spectroscopy} and a previously available line list.}
\label{tbl:comp_dtb}
\centering
\begin{tabular}{cccc@{\hspace*{2\tabcolsep}}cccc c c r}
\hline\hline
$J'$ & $K'$ & $p'$ & $L'$ &
$J''$ & $K''$ & $p''$ & $L''$ &
$F$\tablefootmark{a} & $F$\tablefootmark{b} & $\Delta F$\tablefootmark{c} \\ \hline
3 & 2 & 1 & $o_1$ & 2 & 2 & 1 & $o_1$ &  124964.268 &  124987.573 & -23 \\
3 & 2 & 1 & $e_0$ & 2 & 2 & 1 & $e_0$ &  125018.701 &  125031.220 & -13 \\
3 & 2 & 2 & $o_1$ & 2 & 2 & 2 & $o_1$ &  125031.220 &  125018.714 & 13 \\
4 & 3 & 2 & $e_1$ & 3 & 3 & 2 & $e_1$ &  166297.556 &  166622.635 & -325 \\
4 & 3 & 1 & $e_1$ & 3 & 3 & 1 & $e_1$ &  166297.679 &  166622.635 & -325 \\
4 & 2 & 1 & $o_1$ & 3 & 2 & 1 & $o_1$ &  166608.434 &  166584.647 & 24 \\
4 & 2 & 2 & $o_1$ & 3 & 2 & 2 & $o_1$ &  166775.754 &  166789.815 & -14 \\
5 & 3 & 2 & $e_1$ & 4 & 3 & 2 & $e_1$ &  207868.305 &  208290.307 & -422 \\
5 & 3 & 1 & $e_1$ & 4 & 3 & 1 & $e_1$ &  207868.736 &  208289.062 & -420 \\
6 & 5 & 1 & $e_1$ & 5 & 5 & 1 & $e_1$ &  249400.639 &  249405.569 & -5 \\
6 & 5 & 2 & $e_1$ & 5 & 5 & 2 & $e_1$ &  249400.639 &  249405.569 & -5 \\
6 & 4 & 1 & $e_1$ & 5 & 4 & 1 & $e_1$ &  249429.563 &  249419.799 & 10 \\
6 & 4 & 2 & $e_1$ & 5 & 4 & 2 & $e_1$ &  249429.563 &  249419.799 & 10 \\
6 & 3 & 2 & $e_1$ & 5 & 3 & 2 & $e_1$ &  249436.550 &  249975.994 & -539 \\
6 & 3 & 1 & $e_1$ & 5 & 3 & 1 & $e_1$ &  249437.730 &  249972.691 & -535 \\
7 & 3 & 2 & $e_1$ & 6 & 3 & 2 & $e_1$ &  291001.713 &  291651.893 & -650 \\
7 & 3 & 1 & $e_1$ & 6 & 3 & 1 & $e_1$ &  291004.306 &  291644.451 & -640 \\
7 & 2 & 1 & $o_1$ & 6 & 2 & 1 & $o_1$ &  291477.556 &  291514.936 & -37 \\
7 & 6 & 1 & $o_1$ & 6 & 6 & 1 & $o_1$ &  291501.533 &  291509.826 & -8 \\
7 & 6 & 2 & $o_1$ & 6 & 6 & 2 & $o_1$ &  291501.533 &  291509.826 & -8 \\
7 & 6 & 1 & $e_0$ & 6 & 6 & 1 & $e_0$ &  291509.860 &  291514.936 & -5 \\
7 & 6 & 2 & $e_0$ & 6 & 6 & 2 & $e_0$ &  291509.860 &  291514.936 & -5 \\
8 & 3 & 2 & $e_1$ & 7 & 3 & 2 & $e_1$ &  332563.111 &  333350.925 & -788 \\
8 & 3 & 1 & $e_1$ & 7 & 3 & 1 & $e_1$ &  332568.325 &  333335.082 & -767 \\
8 & 2 & 1 & $e_1$ & 7 & 2 & 1 & $e_1$ &  332719.680 &  332718.625 & 1 \\
8 & 6 & 1 & $e_0$ & 7 & 6 & 1 & $e_0$ &  333151.638 &  333159.463 & -8 \\
8 & 6 & 2 & $e_0$ & 7 & 6 & 2 & $e_0$ &  333151.638 &  333159.463 & -8 \\
8 & 7 & 1 & $o_1$ & 7 & 7 & 1 & $o_1$ &  333159.539 &  333163.863 & -4 \\
8 & 7 & 2 & $o_1$ & 7 & 7 & 2 & $o_1$ &  333159.539 &  333163.863 & -4 \\
8 & 3 & 1 & $o_1$ & 7 & 3 & 1 & $o_1$ &  333335.082 &  333336.082 & -1 \\
8 & 4 & 1 & $e_0$ & 7 & 4 & 1 & $e_0$ &  333354.017 &  333350.239 & 4 \\
8 & 4 & 2 & $e_0$ & 7 & 4 & 2 & $e_0$ &  333354.783 &  333350.239 & 5 \\
\hline
\end{tabular}
\tablefoot{The transition frequencies reported in table~2 of \citet{Mukhopadhyay2016a} are compared to those calculated in this work. Only transitions with a discrepancy larger than $0.5$~MHz appear. Transitions are assigned using the rotational quantum numbers $J,K,p$ \citep{Coudert2021} and the torsional label \citep{SuQuade1989b} of the upper and lower levels.\\
\tablefoottext{a}{The frequency calculated in this work is given in MHz.}\\
\tablefoottext{b}{The frequency reported in \citet{Mukhopadhyay2016a} is given in MHz.}\\
\tablefoottext{c}{The difference between the frequency calculated in this work and that reported in \citet{Mukhopadhyay2016a} is given in MHz.}}
\end{table*}

\subsection{Spectral analysis}
\label{spectral_analysis}
The primary focus of this publication is a one-beam offset position from source B in the SW direction ($0.5\arcsec$ or $70$~au away from the dust peak of B). This is the same position studied in nearly all other ALMA-PILS papers, thereby providing a unique coherent set of abundances for a large number of molecules. This position is favorable for spectral analysis as it displays narrow lines, which reduces blending, and probes the hot dense inner hot corino around source B, while avoiding self-absorption and strong dust absorption on source. Infall signatures have been observed towards source B, which suggests that infall timescales are short and that the hot corino contains freshly thermally desorbed ices \citep{Schoier2002, Zapata2013}. The chosen position likely experiences heating solely from the central protostar, because of its farness from the collimated outflows of IRAS~16293-2422 oriented in NW-SE direction and those less collimated in the E-W direction (fig.~$1$ of \citealt{vanderWiel2019}). The coordinates of this position are RA, Dec (J2000) of 16h32m22.58s, -24d28$\arcmin$32.80$\arcsec$.

Once this single position spectrum is extracted from the continuum-subtracted (based on methodology described in \citealt{Jorgensen2016}) data cube, it is analyzed with custom-made Python scripts. A Doppler shift by $v_{\text{LSR}}=2.7$~km~s$^{-1}$ determined for this position is applied \citep{Jorgensen2011}. As a large number of molecules have already been quantified at this position, it is possible to construct a ``reference spectrum''. This spectrum contains computed emission lines from all molecules detected with the ALMA-PILS survey at this position at their determined excitation temperatures ($T_{\text{ex}}$) and column densities ($N$) with publicly available spectroscopy in either the Cologne Database of Molecular Spectroscopy (CDMS; \citealt{Mueller2001, Mueller2005, Endres2016}) or the JPL catalog \citep{Pickett1998}. The computation is performed under the local thermal equilibrium (LTE) approximation. In many instances, the ``reference spectrum'' overproduces the observed emission, which is a result of many observed lines being optically thick \citep{Jorgensen2018} and requiring special treatment for proper synthetic fitting. Next, the observed spectrum is investigated at all covered rest frequencies of CHD$_{2}$OH lines with the reference spectrum overlaid. In this way, all covered (detected and non-detected) lines of CHD$_{2}$OH are investigated for potential blending with already-identified molecules in this source. This check for blending is done through visual inspection; and all lines that have any level of blending are removed from further synthetic spectrum fitting. Then, optically thick lines of CHD$_{2}$OH are removed from further synthetic spectrum fitting. These are assumed to be lines that have $\tau>0.1$ either at $T_{\text{ex}}=50$ or $300$~K at $N=1.3\times10^{17}$~cm$^{-2}$ (initial rough estimate of the column density of CHD$_{2}$OH). There are in total $554$ lines of CHD$_{2}$OH covered in the observed ALMA-PILS frequency range, of which $525$ have unique rest frequencies. After the removal of blended and optically thick lines, $105$ unique line frequencies remain that are used for synthetic spectral fitting (this includes detected and non-detected lines). Out of $105$ lines, $26$ are predicted to have a peak intensity $>14$~mJy~beam$^{-1}$ (and an integrated intensity $\gtrsim3\sigma$) for the subsequently derived best-fitting $T_{\text{ex}}$ and $N$ (Table~\ref{tbl:N_Tex})\footnote{Within this set of $26$ lines, there are $22$ unique line frequencies. $20$ out of $22$ frequencies correspond to clearly detected emission lines. The remaining two lines are highly excited lines that are difficult to fit with all considered $T_{\text{ex}}$ values, specifically at $343.179$ and $362.018$~GHz (top left and top right panels of Fig.~\ref{fig:CD2HOH_panel_plot_150K_300K_non_det}).}. Non-blended non-detected lines are of course also constraining the synthetic spectral fitting.

Finally, synthetic spectral fitting for CHD$_{2}$OH is carried out with a newly developed routine based on the Markov chain Monte Carlo (MCMC) Python package \textsc{emcee} \citep{ForemanMackey2013}\footnote{\url{https://emcee.readthedocs.io/en/stable/}}. The code computes an LTE synthetic spectrum based on several input parameters. Several parameters are fixed: beam size of $0.5\arcsec$, source size of $0.5\arcsec$, and line full width half-maximum (FWHM) of $1$~km~s$^{-1}$. The parameter space of $T_{\text{ex}}$ and $N$ is explored by walkers that take a certain number of steps. Two runs of MCMC are executed. The first has $300$ walkers for each parameter with each walker taking $250$ steps, which are initiated at values distributed in a Gaussian around the medians of ``sensible'' $T_{\text{ex}}$ and $N$ ranges, specifically: $T_{\text{ex}}\in[50,400]$~K and $N\in[10^{13},10^{18}]$~cm$^{-2}$ (the log-prior function is set to infinity outside of these ranges). The second has $300$ walkers for each parameter with each taking $1000$ steps, but now initiated in a small region around the best-fitting $T_{\text{ex}}$ and $N$ of the first run. The best-fitting $T_{\text{ex}}$ and $N$ of the second run are considered the final best-fitting model parameters to the data. In both runs, the first $10\%$ of the steps are discarded as this corresponds to the burn-in phase of the computation (that this is sufficient is subsequently double checked with the autocorrelation times). Both runs are also thinned by $0.3\%$ of the steps. The best-fitting values are considered the 50th percentiles of the samples in the marginalized distributions. The errors computed by the routine are considered to be the differences of the 50th percentile to the 16th and 84th percentiles (i.e., $1\sigma$ errors if the values have a Gaussian distribution centered on the mean). These errors are very small for well-converged runs and do not take all origins of uncertainty in observations into account (see a continuation of this discussion in Sect.~\ref{line_profiles}). The walkers undertake the default ``stretch moves'' of \textsc{emcee} \citep{GoodmanWeare2010}. The negative likelihood function being minimized is:
\begin{equation}
\label{min_eq}
-\frac{1}{2} \times \sum_{\nu_{line}} \sum_{\nu_{line}-0.5\times\text{FWHM}}^{\nu_{line}+0.5\times\text{FWHM}} \left( I_{\text{observed}}(\nu)-I_{\text{model}}(\nu) \right)^{2},
\end{equation}
where the outer summation considers rest frequencies ($\nu_{line}$) of all optically thin, non-blended (detected and non-detected) lines covered in the observed ALMA-PILS frequency range. The inner summation means that the model is computed and compared to observations around each line in a range of $\pm\text{FWHM}/2$ around its respective rest frequency. Eq.~\ref{min_eq} implies that all optically thin, non-blended (detected and non-detected) lines are being considered simultaneously. The output of the second MCMC run is checked for convergence and high values of acceptance ratios. For the second MCMC run of the CHD$_{2}$OH computation, the mean acceptance fraction of the $300$ walkers is $70\%$ and the quality of the convergence is illustrated in the corner plot shown in Fig.~\ref{fig:CD2HOH_v1_corner2} for I16293B. For I16293A, the second MCMC run of the CHD$_{2}$OH computation has the mean acceptance fraction of the $300$ walkers also at $70\%$ and the quality of the convergence is illustrated in the corner plot shown in Fig.~\ref{fig:CD2HOH_v1_corner2_A}.

Fundamentally, MCMC is used to compare a model to data. It is inherently a sampler \citep{ForemanMackey2013}\footnote{Also, useful, e.g., \url{https://prappleizer.github.io/Tutorials/MCMC/MCMC_Tutorial.html} and \url{https://jellis18.github.io/post/2018-01-02-mcmc-part1/}.}. Consequently, the fit is only as good as the postulated model that is thought to match the observed data. In the case of this paper, this is the LTE synthetic spectrum model. It is thought that this is a good approximation, because the densities are high and molecules are expected to be thermalized (section~$5.1$ of \citealt{Jorgensen2016}). The code behind the LTE model is based on the formalisms used in CASSIS. The partition function is linearly interpolated in log-space between the values corresponding to the two temperatures closest to the $T_{\text{ex}}$ being computed. The code accounts for optical depth by considering the Sobolev or the large velocity gradient (LVG) approximation: $N_{\text{up; op. thick}} = N_{\text{up; op. thin}} \times \beta$, where $\beta = \left( 1-e^{-\tau} \right) / \tau$ (\citealt{Sobolev1960, Castor1970, Elitzur1992}; as in RADEX, p.42-44 of \citealt{vanderTak2007}). However, it does not properly account for flat-topped line profiles of optically thick lines, therefore being most-suitable for the analysis of optically thin lines. With both beam and source distributions being Gaussian, the dilution of the observed emission is accounted for by the beam-filling factor: $\eta_{\text{BF}} = \text{source size}^{2} / \left( \text{source size}^{2} + \text{beam size}^{2} \right) = 0.5$ \citep{Drozdovskaya2018}.

The secondary focus of this publication is a (slightly more than) one-beam offset position from source A in the NE direction ($0.6\arcsec$ or $85$~au away from the dust peak of A). This is the same position studied in all other ALMA-PILS papers that analyzed source A with coordinates RA, Dec (J2000) of 16h32m22.90s, -24d28$\arcmin$36.20$\arcsec$ and $v_{\text{LSR}}=0.8$~km~s$^{-1}$. The same spectral analysis methodology is applied to this position as for the offset from source B position. This excludes the same lines based on blending and optical thickness grounds and uses the same $105$ unique line frequencies for synthetic spectral fitting. All other input parameters are kept the same; only FWHM is changed to $2.2$~km~s$^{-1}$ as deduced from previous analysis of this position \citep{Manigand2020}.


\section{Results}
\label{results}
\subsection{Line profiles}
\label{line_profiles}
\begin{figure}
	\centering
	\includegraphics[width=\hsize]{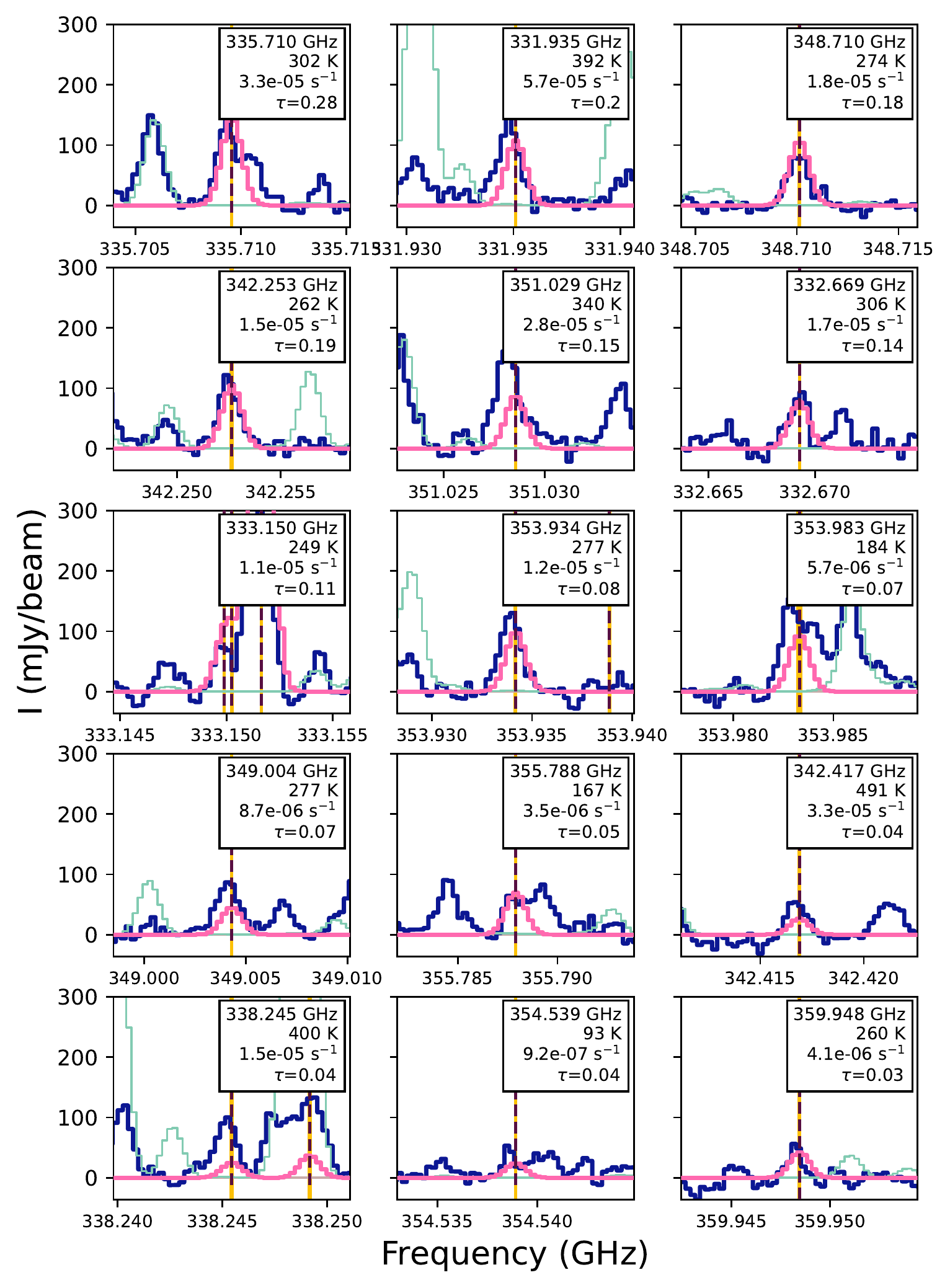}
	\caption{Selection of some of the strongest $15$ lines of CHD$_{2}$OH out of the $105$ used for synthetic spectral fitting. The observed spectrum is in dark blue, the ``reference'' spectrum is in turquoise, and the best-fitting synthetic spectrum is in pink. The rest frequency, $E_{\text{up}}$ (K), $A_{ij}$ (s$^{-1}$), and optical depth (for the best-fitting parameters) are shown in the right corner of each panel. The rest frequency is indicated with a vertical dashed line, and the filled yellow region corresponds to the uncertainty on that line frequency. Note that the lines at $353.934$, $353.983$, $355.788$, and $359.948$~GHz are in fact overlapping double transitions.}
	\label{fig:CD2HOH_panel_plot}
\end{figure}

A selection of some of the strongest $15$ lines of CHD$_{2}$OH out of the $105$ used for synthetic spectral fitting is shown in Fig.~\ref{fig:CD2HOH_panel_plot} for source B (and in Fig.~\ref{fig:CD2HOH_panel_plot_A} for source A). Since these have already been filtered for blended and optically thick lines (Sect.~\ref{spectral_analysis}), all the shown lines are not blending with any of the known molecules of I16293B (there may still be blending with unknown species) and are also mostly optically thin ($\tau < 0.1$ cut-off holds for the initially guessed $N=1.3\times10^{17}$~cm$^{-2}$ and $T_{\text{ex}} = 50$~ and $300$~K; Sect.~\ref{spectral_analysis}). The best-fitting synthetic spectrum is shown in pink in Fig.~\ref{fig:CD2HOH_panel_plot} and corresponds to $T_{\text{ex}}=90.0\pm9$~K and $N=(1.78\pm0.36)\times10^{17}$~cm$^{-2}$. The errors that are derived during the MCMC computation are very small, because they are measuring the goodness of fit of the assumed model to the data. However, the model assumptions (e.g., LTE with a single $T_{\text{ex}}$) have much higher uncertainties and so do the data themselves (calibration uncertainty of $5\%$). These uncertainties are not taken into account in the MCMC computation. It is hard to quantify the exact uncertainty on the parameters, consequently, pessimistic errors of $10\%$ and $20\%$ on $T_{\text{ex}}$ and $N$, respectively, are adopted in order to ensure that the uncertainties are not underestimated \citep{Jorgensen2018}. The observed spectrum contains much stronger lines of CHD$_{2}$OH, which have been excluded from fitting due to their high $\tau$ values. Another selection of some of the strongest $15$ lines of CHD$_{2}$OH observed (all optically thick) is shown in Fig.~\ref{fig:CD2HOH_panel_plot_top15}. Figs.~\ref{fig:CD2HOH_panel_plot} and~\ref{fig:CD2HOH_panel_plot_top15} support the clear detection of CHD$_{2}$OH towards I16293B with consistent narrow ($\sim1$~km~s$^{-1}$) lines and demonstrate the quality of the synthetic spectral fitting.

The excitation temperature for IRAS16293B deduced by the MCMC fitting is lower than the $\approx$300~K previously inferred for CH$_{3}^{18}$OH \citep{Jorgensen2016} and CH$_{3}$OD \citep{Jorgensen2018}. For example, rotational diagram fits to the CH$_{3}$OD transitions showed a best-fitting $321\pm31$~K and could not support a lower excitation temperature of $125$~K (fig.~$1$ of \citealt{Jorgensen2018}). To investigate this further, MCMC models for fixed temperatures of $150$ and $300$~K were ran while varying the column density. The results of these models are shown in Figs.~\ref{fig:CD2HOH_panel_plot_150K_300K} and~\ref{fig:CD2HOH_panel_plot_150K_300K_non_det} in Appendix~\ref{additional_figures} for the $15$ strongest lines (included in the synthetic spectral fitting) and for a set of $15$ high excitation transitions, respectively. While the former do not show significant differences for the different excitation temperatures, the latter do: a couple of high excitation transitions ($E_{\text{up}}\approx500-700$~K) are predicted at the $50-100$~mJy/beam level, but are not seen in the observed spectra. With the lower temperature of the $90$~K model, less emission is seen for those lines; although, there are still examples of over-predicted emission for some. In the end, this set of lines drives the best-fitting $T_{\text{ex}}$ to the lower values: whether this is reflecting actual differences between the methanol isotopologs or rather uncertainties in the spectroscopy for these transitions remains unclear. No matter, the different MCMC models allow us to quantify the uncertainties on $N$ as a result of changing $T_{\text{ex}}$. For the $T_{\text{ex}}=300$~K model the best-fitting column density is $N=(6.62\pm1.32)\times10^{16}$~cm$^{-2}$, while for the $T_{\text{ex}}=150$~K model the best-fitting column density is $N=(1.07\pm0.21)\times10^{17}$~cm$^{-2}$ from the MCMC computation. Consequently, the lower $N$ values for higher $T_{\text{ex}}$ result in lower D/H ratios derived for CHD$_{2}$OH (by at most a factor of $1.6$); however, the conclusions of this paper are not changed even when considering this hypothetical broader range of D/H ratios (Table~\ref{tbl:N_Tex}). The excitation temperature for IRAS16293A deduced by the MCMC fitting is consistent with previously published values for this source \citep{Calcutt2018a, Manigand2020}.

\begin{table*}
\caption{Best-fitting column densities, excitation temperatures, and D/H ratios (with statistical correction) of CHD$_{2}$OH towards IRAS~16293-2422~B and~A.}
\label{tbl:N_Tex}
\centering
\begin{tabular}{l l l}
\hline\hline
Best-fitting parameters                           & I16293B                             & I16293A\T\B\\
\hline
$T_{\text{ex}}$ (K)                               & $90.0\pm9$                        & $127\pm13$\T\\
$N$ (cm$^{-2}$)                                   & $(1.78\pm0.36)\times10^{17}$       & $(2.22\pm0.44)\times10^{17}$\\
D/H of CH$_{3}$OH (w/ stat. corr.)                & $(7.7\pm1.2)\times10^{-2}$          & $(7.5\pm1.1)\times10^{-2}$\B\\
\hline
$N$ (cm$^{-2}$) at $T_{\text{ex}}=150$~K          & $(1.07\pm0.21)\times10^{17}$        & $\cdots$\T\\
D/H of CH$_{3}$OH (w/ stat. corr.)                & $(6.0\pm0.9)\times10^{-2}$          & $\cdots$\B\\
\hline
$N$ (cm$^{-2}$) at $T_{\text{ex}}=300$~K          & $(6.62\pm1.32)\times10^{16}$        & $\cdots$\T\\
D/H of CH$_{3}$OH (w/ stat. corr.)                & $(4.7\pm0.7)\times10^{-2}$          & $\cdots$\B\\
\hline
$N($CH$_{3}$OH$)$ (cm$^{-2}$)                     & $(1.0\pm0.2)\times10^{19}$ & $(1.3\pm0.4)\times10^{19}$\T\\
Reference                                         & \citet{Jorgensen2018}      & \citet{Manigand2020}\B\\
\hline
\end{tabular}
\tablefoot{Top two rows are the best-fitting column density and excitation temperature of CHD$_{2}$OH obtained during simultaneous fitting of these two parameters. The associated D/H ratio with statistical correction is in the third row (Appendix~\ref{statistical_corrections}). The cases of constant $T_{\text{ex}}=150$ and $300$~K have also been computed with the corresponding derived column densities and D/H ratios in rows four and five, and six and seven, respectively. The second column contains the values for the one-beam offset position in the SW direction from IRAS~16293-2422~B ($0.5\arcsec$ or $70$~au away from the dust peak of B). The third column contains the values for the one-beam offset position in the NE direction from IRAS~16293-2422~A ($0.6\arcsec$ or $85$~au away from the dust peak of A). The last two rows give the methanol column densities adopted for the calculations of D/H ratios and the corresponding references. The stated errors on $N$ and $T_{\text{ex}}$ are assumed $20\%$ and $10\%$ errors, respectively. The errors derived with the MCMC routine in this work are much smaller, refer to Sects.~\ref{spectral_analysis} and~\ref{line_profiles} for a discussion about that. For $N($CH$_{3}$OH$)$ in I16293A, the error is $30\%$ as derived in \citet{Manigand2020}.}
\end{table*}

\subsection{Integrated intensity maps}
\label{integrated_intensity_maps}
\begin{figure}
	\centering
	\begin{subfigure}[b]{0.45\textwidth}
		\includegraphics[width=\hsize]{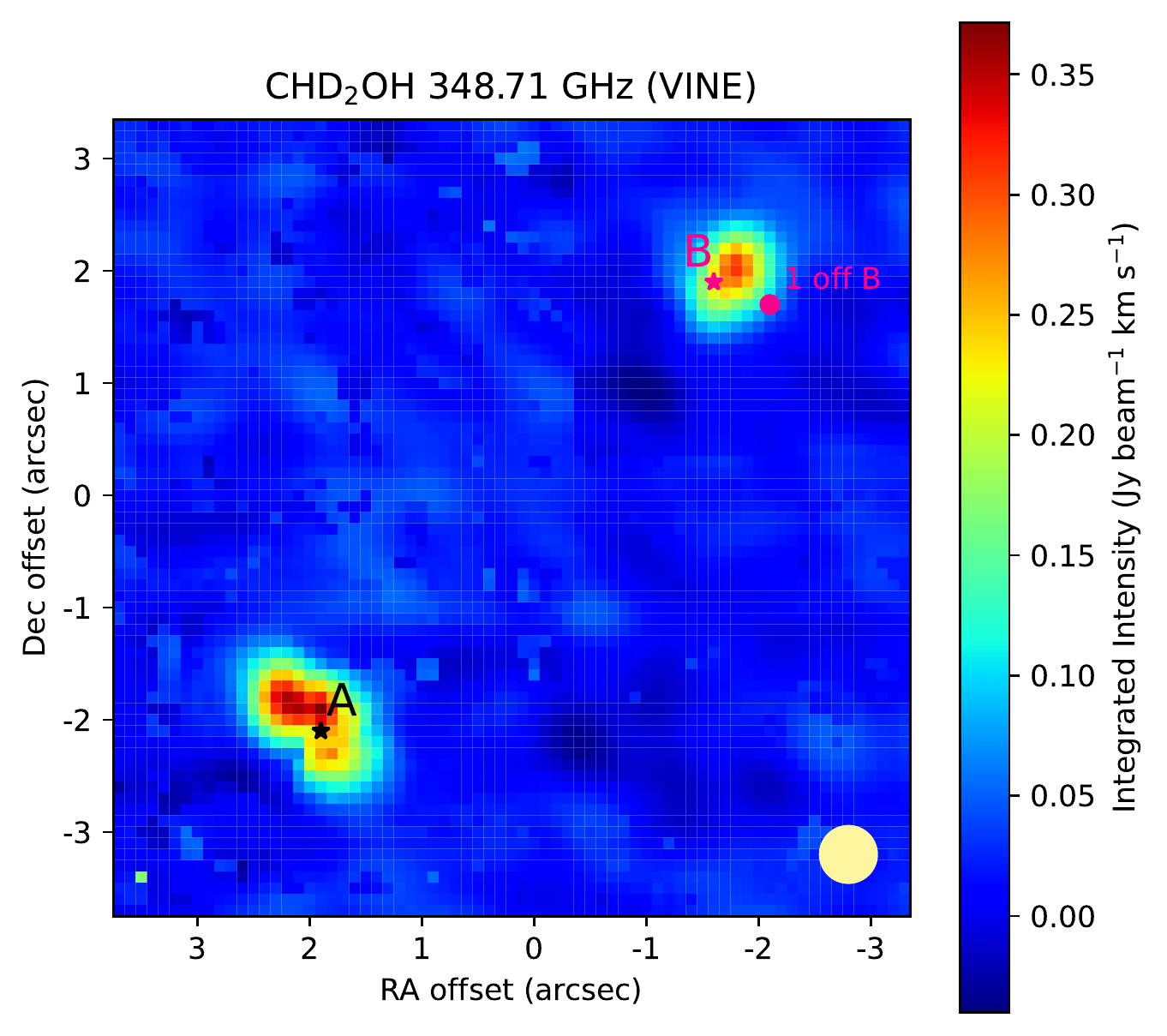}
		\caption{(nearly) optically thin $348.710$~GHz line}
		\label{fig:CD2HOH_348.7101234_vine_multi_paper}
	\end{subfigure}
		\begin{subfigure}[b]{0.45\textwidth}
		\includegraphics[width=\hsize]{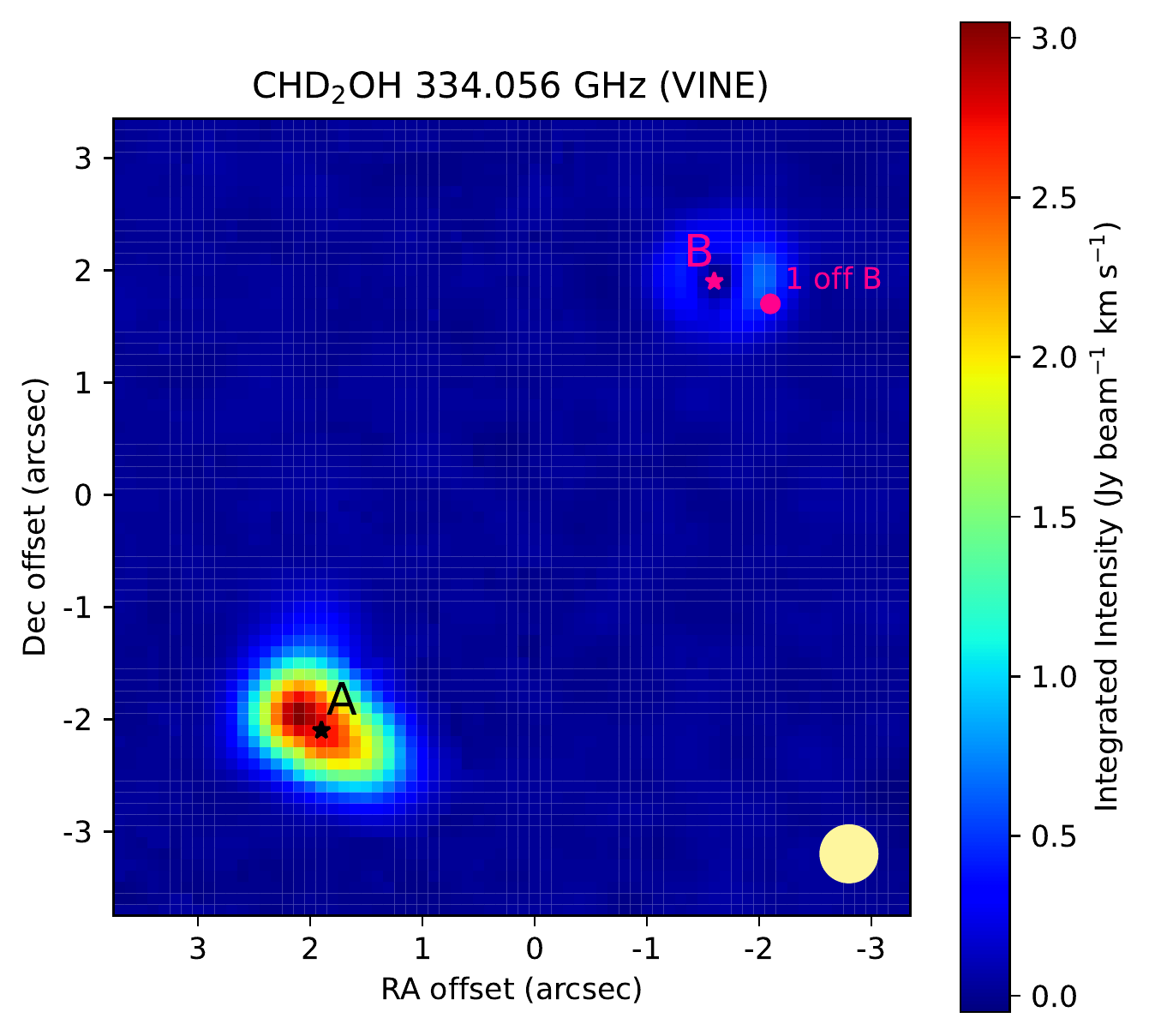}
		\caption{optically thick $334.056$~GHz line}
		\label{fig:CD2HOH_334.056369_vine_multi_paper}
	\end{subfigure}
	\caption{Integrated intensity maps of CHD$_{2}$OH (moment 0 maps). The binary sources A and B are labeled, as well as the one-beam offset position in the SW direction from source B that is central to this work. The origin of the axes corresponds to RA, Dec (J2000) of 16h32m22.72s, -24d28$\arcmin$34.3$\arcsec$. The $0.5\arcsec$ beam size is indicated in the bottom right by a yellow circle.}
\end{figure}

The spectral line richness of IRAS~16293-2422 complicates the creation of moment 0 maps. When an integration across a wide velocity range is performed (in order to encompass all the emission in the line of interest), it becomes increasingly easy to also erroneously integrate over emission from some adjacent line of another molecule. This is particularly exacerbated near source A, as it has much broader lines than source B: FWHM$_{\text{A}}\approx2.2$~km~s$^{-1}$, which is a factor of 2 higher than FWHM$_{\text{B}}\approx1$~km~s$^{-1}$ at the most favorable position near source A in terms of line width (some positions near source A can reach FWHMs of $\sim5$~km~s$^{-1}$). This means that for IRAS~16293-2422, meaningful moment 0 maps can only be created for very few, highly isolated lines. The (nearly) optically thin $348.710$~GHz line of CHD$_{2}$OH is highly promising in this regard (the line profile towards the one-beam offset position is shown in Fig.~\ref{fig:CD2HOH_panel_plot}: top row, right column). For the best-fitting parameters of the one-beam offset position from B, this line has $\tau=0.18$. Fig.~\ref{fig:CD2HOH_348.7101234_vine_multi_paper} shows the integrated intensity map of this line. Around source A it is still hard to compute an accurate integrated intensity and therefore, the shown map should be interpreted with caution (see Appendix~\ref{integrated_intensity_maps_details}). Here, the integration is performed over $25$ spectral bins, corresponding to a $\sim5$~km~s$^{-1}$ range, which is mostly sufficient to encompass the entire line profile near both sources. The rest frequency of the line is characterized well with a single value only in the vicinity of source B. Around source A, a steep velocity gradient exists \citep{Pineda2012, Maureira2020a}. Consequently, the rest frequency of this line in every pixel is found by first determining the rest frequency of the bright methanol transition at $337.519$~GHz, thereby determining a velocity map of IRAS~16293-2422 (fig.~$4$ of \citealt{Calcutt2018a}). This method was developed by \citet{Calcutt2018a} and produces so-called Velocity-corrected INtegrated emission (VINE) maps. 

Fig.~\ref{fig:CD2HOH_348.7101234_vine_multi_paper} illustrates that CHD$_{2}$OH emission is compact towards A and B, but that a more detailed analysis of the spatial distribution would require data of higher spatial resolution (beam size $<0.5\arcsec$). Fig.~\ref{fig:CD2HOH_334.056369_vine_multi_paper} shows the integrated intensity map of the optically thick $334.056$~GHz line of CHD$_{2}$OH. The line profile towards the one-beam offset position is shown in Fig.~\ref{fig:CD2HOH_panel_plot_top15} (bottom row, right column). The peak integrated intensity in this map is roughly an order of magnitude higher than in Fig.~\ref{fig:CD2HOH_348.7101234_vine_multi_paper}, as this is a much brighter line of CHD$_{2}$OH. For the best-fitting parameters of the one-beam offset position from B, this line has $\tau=7.91$ and the combined effects of line optical depth and on-source dust optical depth are clearly visible in Fig.~\ref{fig:CD2HOH_334.056369_vine_multi_paper} in the form of an emission ring around source B. Both Figs.~\ref{fig:CD2HOH_348.7101234_vine_multi_paper} and~\ref{fig:CD2HOH_334.056369_vine_multi_paper} show an asymmetry near source B, which is seen for most molecules in the hot corino. The origin of this asymmetry remains to be conclusively unraveled (e.g., \citealt{Zamponi2021}).

\section{Discussion}
\label{discussion}

\begin{figure*}
	\centering
	\includegraphics[width=\hsize]{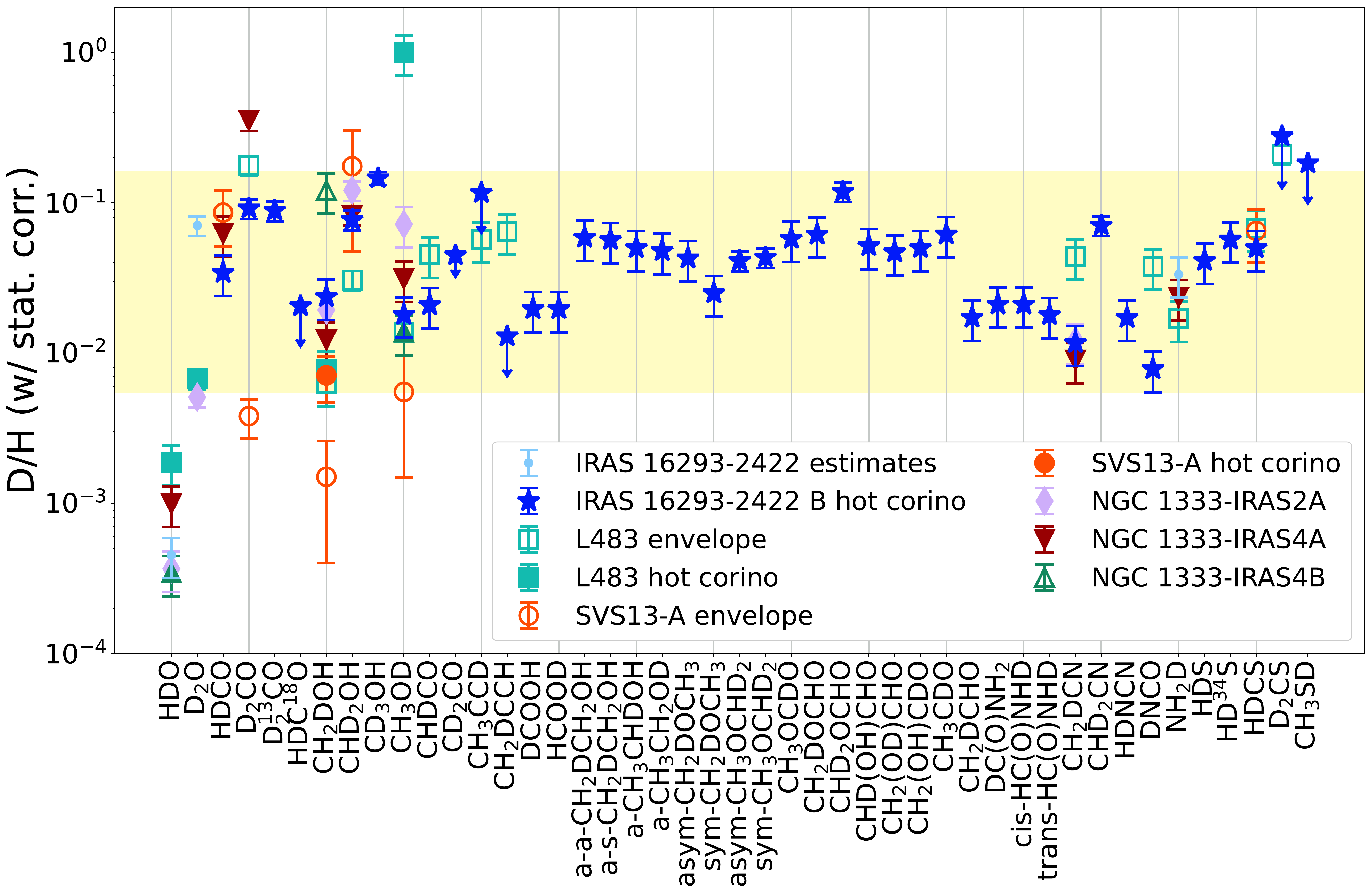}
	\caption{D/H ratio as measured in all the D-bearing molecules detected in the hot corino of IRAS~16293-2422~B, including the respective statistical corrections (Appendix~\ref{statistical_corrections}). Available D/H ratios measured in other low-mass star-forming systems are also shown, see the main text for details.}
	\label{fig:DH_plot_protostars}
\end{figure*}

\begin{figure}
	\centering
	\includegraphics[width=\hsize]{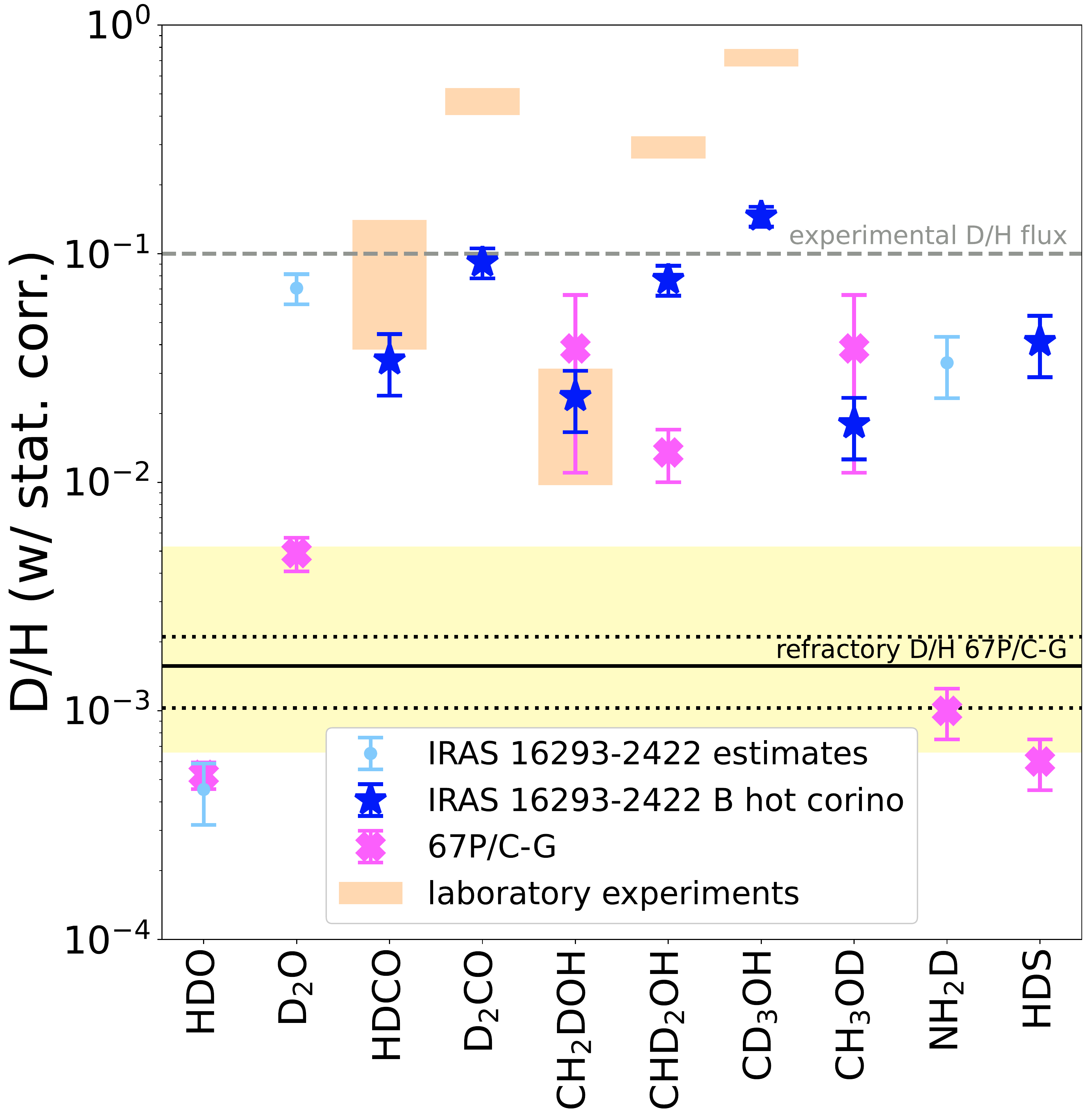}
	\caption{D/H ratio as measured in the hot corino of IRAS~16293-2422~B, in the volatiles and dust of comet 67P/Churyumov–Gerasimenko, and in laboratory experiments. The respective statistical corrections (Appendix~\ref{statistical_corrections}) are accounted for in all data points. The dashed gray line is the experimental D/H input flux. The solid black line is the mean value of the refractory carbonaceous component with black dashed lines indicating the errors on this mean and the yellow region indicating the full range of values covered by the individual $25$ particle measurements. See the main text for details and references.}
	\label{fig:DH_plot_exp}
\end{figure}

\subsection{Deuteration in IRAS~16293-2422}
\label{deuteration}
A large number of molecules have been detected in IRAS~16293-2422, including many deuterated species (appendix~C of \citealt{Drozdovskaya2019}). Fig.~\ref{fig:DH_plot_protostars} shows the D/H ratio as measured in all the D-bearing molecules detected in the hot corino of I16293B, including the respective statistical corrections (Appendix~\ref{statistical_corrections}). All the data points (except for water and ammonia) of I16293B stem from the ALMA-PILS survey as determined at the same one-beam offset position in all the respective publications. The plotted HDO/H$_{2}$O ratio is an estimate on disk-scales towards source A \citep{Persson2013}. The D$_{2}$O/H$_{2}$O ratio is an estimate on scales of the circumbinary envelope encompassing sources A and B together \citep{Coutens2013a}. The plotted NH$_{2}$D/NH$_{3}$ ratio is a circumbinary estimate \citep{vanDishoeck1995}. Upper limits determined for D$_{2}$CS \citep{Drozdovskaya2018}, CH$_{3}$SD \citep{Zakharenko2019a}, CD$_{2}$CO \citep{Jorgensen2018}, HDC$^{18}$O \citep{Persson2018}, and the two mono-deuterated isotopologs of propyne (\citealt{Calcutt2019}; although, \citealt{Caux2011} did secure a detection of CH$_{2}$DCCH as part of ``The IRAS16293-2422 Millimeter And Submillimeter Spectral Survey'', TIMASSS, most likely from the circumbinary envelope) are also depicted for completeness.

In Fig.~\ref{fig:DH_plot_protostars}, the error bars on all the values of I16293B are a result of the $20\%$ error on each $N$, which gives an error of $30\%$ on the ratio of two column densities, and a result of statistical error propagation based on the respective statistical corrections. The hot corino D/H ratios of all available molecules (including error bars) lie in the $5.5\times10^{-3} - 1.6\times10^{-1}$ range (excluding estimated ratios of water and ammonia, and upper limits). No apparent trend of the D/H ratio with functional group is seen, as was already realized by \citet{Jorgensen2018}. Fig.~\ref{fig:DH_plot_protostars} includes molecules with deuteration in the hydroxyl group (HDO, D$_{2}$O, CH$_{3}$OD, HCOOD, a-CH$_{3}$CH$_{2}$OD), the aldehyde group (CH$_{2}$(OD)CHO),  the amino group (NH$_{2}$D, cis- and trans-HC(O)NHD, HCNCN, DNCO), the thiol group (HDS, HD$^{34}$S), and all others in the methylene (CH$_{2}-$) or methyl (CH$_{3}-$) groups. No apparent trend is seen for different families of molecules either, i.e., CHO-, N-, S-bearing, when considering all D-bearing molecules. A tentative trend is seen when considering solely mono-deuterated species: D/H ratios seem to be lower for N-bearing molecules and O-bearing species that contain less than 6 atoms in comparison to S-bearing molecules and O-bearing species with more than 6 atoms. D/H ratios lower by a factor of $2-4$ were previously deduced for small O-bearing molecules than those for large O-bearing species \citep{Jorgensen2018}. This may be related to the differences in formation timescales of small and large molecules as proposed in \citet{Coutens2020}. Mono- and di-deuterated molecules of I16293B are shown separately in Figs.~\ref{fig:DH_plot_protostars_mono} and~\ref{fig:DH_plot_protostars_di}.

What does persist is the trend of di-deuterated species having a higher D/H ratio than their mono-deuterated counterparts. The fractionation levels of mono-deuterated formaldehyde, methanol, methyl formate, dimethyl ether, and methyl cyanide are:
\begin{align}
&\left(\frac{\text{D}}{\text{H}}\right)_{\text{HDCO}} &=& 2 \times \frac{\text{D}_{2}\text{CO}}{\text{HDCO}} &\approx& 24.6 \pm 7.4 \%,\\
&\left(\frac{\text{D}}{\text{H}}\right)_{\text{CH}_{2}\text{DOH}} &=& \frac{\text{CHD}_{2}\text{OH}}{\text{CH}_{2}\text{DOH}} &\approx& 25.1 \pm 7.5 \%,\\
&\left(\frac{\text{D}}{\text{H}}\right)_{\text{CH}_{2}\text{DOCHO}} &=& \frac{\text{CHD}_{2}\text{OCHO}}{\text{CH}_{2}\text{DOCHO}} &\approx& 22.9 \pm 6.9 \%,\\
&\left(\frac{\text{D}}{\text{H}}\right)_{\text{CH}_{3}\text{OCH}_{2}\text{D}} &=& \frac{\text{CH}_{3}\text{OCHD}_{2}}{\text{CH}_{3}\text{OCH}_{2}\text{D}} &\approx& 16.6 \pm 5.0 \%,\\
&\left(\frac{\text{D}}{\text{H}}\right)_{\text{CH}_{2}\text{DCN}} &=& \frac{\text{CHD}_{2}\text{CN}}{\text{CH}_{2}\text{DCN}} &\approx& 14.3 \pm 4.3 \%.
\end{align}
Above, the sym- and asym- variants of mono- and di-deuterated dimethyl ether have been summed (as in \citealt{Richard2021}). Within error bars, all five values agree, implying the same level of deuteration in all di-deuterated species relative to their mono-deuterated variants. The numbers suggest that roughly one in every 4 mono-deuterated molecules of the aforementioned molecules will undergo a second deuteration. However, di-deuterated molecular D/H ratios are only $6$ values out of $36$ (excluding HDO, D$_{2}$O, NH$_{2}$D, and upper limits). Di-deuterated molecular D/H ratios span $10-33\%$ with a mean of $20.7\pm6.2\%$. The D$_{2}$O/HDO ratio is estimated at $0.004$ and $0.063$ for the hot corino and outer envelope, respectively, which would correspond to a $\left(\frac{\text{D}}{\text{H}}\right)_{\text{HDO}}$ of $0.8\%$ and $12.6\%$, respectively \citep{Coutens2013a}. Consequently, D$_{2}$O may be associated with a factor of $\sim2$ lower D/H ratio in comparison to other di-deuterated molecules, at least on the large circumbinary scales of those data.

Even within such a consistent spectral analysis as was carried out based on ALMA-PILS, a certain degree of ambiguity remains when it comes to determining column densities of some of the main isotopologs. These can be derived from their minor carbon or oxygen isotopologs and an assumed local ISM $^{12}$C/$^{13}$C or $^{16}$O/$^{18}$O ratio, or based on a vibrationally excited state, or from the analysis of optically thin lines alone of the main isotopologs themselves. In \citet{Jorgensen2016, Jorgensen2018}, it was already suspected that some species may be associated with a lower $^{12}$C/$^{13}$C ratio than the canonical ISM value. If the applied $^{12}$C/$^{13}$C ratio is too high, then the D/H ratios shown in Fig.~\ref{fig:DH_plot_protostars} would be higher for the species in question. A key species to constrain the column density of would be HD$^{13}$CO, as it would allow the determination of $\left(\frac{\text{D}}{\text{H}}\right)_{\text{HD}^{13}\text{CO}} = 2 \times \frac{\text{D}_{2}^{13}\text{CO}}{\text{HD}^{13}\text{CO}}$. This would test whether the level of deuteration is independent of the $^{12}$C/$^{13}$C ratio. The lines of HD$^{13}$CO covered in the frequency range of the ALMA-PILS were too weak for detection \citep{Persson2018}. More deuterated species have been detected towards I16293B than those displayed in Fig.~\ref{fig:DH_plot_protostars}. All the omitted species are thought to trace the large-scale circumbinary envelope or even the larger surrounding cloud (Appendix~\ref{excluded_deuterated_species}). The D/H ratios measured in I16293A are very similar to those in I16293B on a molecule by molecule basis, as already reported in \citet{Manigand2020}. This high degree of agreement (Fig.~\ref{fig:DH_plot_protostars_A}) supports the key role of the birth cloud in setting the chemical composition of the material in the protostellar systems that form therein.

\subsection{Comparison to other low-mass protostars with CHD$_{2}$OH detections}
\label{lm_protostars}
CHD$_{2}$OH was detected for the first time by \citet{Parise2002} with the IRAM~30~m single-dish telescope in IRAS~16293-2422. Due to the low spatial resolution of the data, it was not possible to disentangle the emitting components. The ALMA-PILS data analyzed in this work are now clearly showing that both sources A and B are emitting in CHD$_{2}$OH from within their respective hot corino regions; unfortunately, A1 and A2 cannot be separated at the present spatial resolution. This molecule has also been detected in the low-mass Class I SVS13-A system \citep{Bianchi2017a}, in the cold envelope of the Class 0 L483 source \citep{Agundez2019}, and in the hot corinos of NGC~1333-IRAS2A, -IRAS4A, and -IRAS4B \citep{Parise2006, Taquet2019}. The uncertainties on CHD$_{2}$OH column densities may be elevated in comparison to that derived in this work in IRAS~16293-2422 due to the significant uncertainties in the past non-public CHD$_{2}$OH spectroscopy and poor rigid-rotor assumptions for its partition function (Sect.~\ref{spectroscopy}). For example, for IRAS~16293-2422 \citet{Parise2002} reported CHD$_{2}$OH/CH$_{3}$OH$=0.2\pm0.1$; however, based on the values in Table~\ref{tbl:N_Tex} this ratio is $1.78\pm0.5\times10^{-2}$ and $1.71\pm0.6\times10^{-2}$ for source B and A, respectively. The large one order of magnitude discrepancy is likely due to less accurate old spectroscopy and underestimates of the CH$_{3}$OH column density stemming from its optically thick lines. It would be fruitful to revisit the CHD$_{2}$OH analysis in these others five sources as is being done in this work for IRAS~16293-2422. Fig.~\ref{fig:DH_plot_protostars} includes deuterated molecules that are common to I16293B and any of these other five sources.

Details of past observations of other low-mass protostellar regions that are being compared to I16293 are presented in Appendix~\ref{additional_figures}. Several aspects of concern must be acknowledged when carrying out such comparisons with observational campaigns stemming from other facilities and based on different assumptions. These include scarcity of analyzed lines (notably of CHD$_{2}$OH) down to a level of just $1-2$ lines, assumed isotopic ratios ($^{12}$C/$^{13}$C), and different treatments of potentially optically thick lines of the main isotopologs. The cumulative uncertainties arising from these aspects are likely partially responsible for the scatter in the D/H ratios in Fig.~\ref{fig:DH_plot_protostars} between sources. However, Fig.~\ref{fig:DH_plot_protostars} does not yield enough evidence to support I16293B being abnormally deuterium-rich among low-mass protostars. Systematic differences in the D/H ratios of several molecules are not seen when comparing it to L483, SVS13-A, NGC1333-IRAS2A, -IRAS4A, and -IRAS4B. Neither are any concrete differences seen across these early evolutionary stages, when comparing these six Class 0 and I sources. I16293B does have a much larger number of D-bearing molecule detections in comparison to other low-mass protostars. Likely, this is an observational bias owing to the small distance to this source, its optimal positioning in the sky resulting in narrow line widths, more thorough and frequent studies of the source by many observational projects and spectral surveys. By now, IRAS~16293-2422 has been studied with an exceptionally wide spectral coverage, which increases the chances of detecting many lines from numerous molecules.

For the limited detections of di-deuterated molecules in these other sources, the trend of D/H ratios of di-deuterated molecules relative to their mono-deuterated counterparts being higher than the D/H ratios of mono-deuterated isotopologs relative to the non-deuterated species appears to persist (Fig.~\ref{fig:DH_plot_protostars}). L483 and NGC1333-IRAS4B are the only non-binary protostars considered in this comparison (Appendix~\ref{additional_figures}). The binarity of the other sources is only spatially resolved for the case of I16293A and I16293B. The D/H ratios of D-bearing molecules do not show any trends with the degree of multiplicity or separation in binaries within this small set of sources. There appears to be a difference in deuterium fractionation between low- and high-mass star-forming regions \citep{Jorgensen2020}, in particular when it comes to deuteration of methanol \citep{Taquet2019} or later stages of high-mass protostellar evolution \citep{Zahorecz2021}. The study of di-deuterated molecules should be expanded to high-mass sources in future work.

\subsection{Comparison to experimental data}
\label{exp_theo}
Fig.~\ref{fig:DH_plot_exp} shows the D/H ratio of formaldehyde and methanol in the hot corino of I16293B in comparison to experimental results of \citet{Nagaoka2005, Nagaoka2007, Hidaka2009}. The results stem from an extensive series of experiments performed on amorphous solid water (ASW) at $10-20$~K (the experimental values stem from N. Watanabe priv. comm.; this can be considered an update of fig.~$5$ of \citealt{Nagaoka2006}). During the experiments, cold H and D atoms are deposited (separately or simultaneously) onto either solid CO or formaldehyde ice (H$_{2}$CO or D$_{2}$CO) with a thickness of $\sim$nm with ASW underneath (at least for the experiments with formaldehyde). Different combinations of reactants were explored in the series of experiments to accurately pin down the chemical network at hand. Sample composition is monitored with reflection-absorption spectroscopy directly in the solid phase. Under laboratory conditions, the D/H ratios of multiply deuterated isotopologs are higher than those of mono-deuterated molecules as observed towards I16293B. Consequently, this observational trend cannot be attributed to multiply deuterated molecules forming under conditions with more D atoms, because the experimental D/H flux is constant. Rather, it seems that this is a natural consequence of H-D substitution reactions on grain-surfaces at cold temperatures. Successive hydrogen abstractions seem to be more likely once the first deuteration of a molecule has occurred. D/H ratios of HDCO and CH$_{2}$DOH match between observations and experiments even though the experimental D/H flux of $0.1$ is much higher than $\sim10^{5}$ that is available in the ISM. The D/H ratios of multiply deuterated species (D$_{2}$CO, CHD$_{2}$OH, and CD$_{3}$OH) are higher than the observed values, which could be a sign of an over-availability of D atoms for the synthesis of multiply deuterated isotopologs in the experiments (i.e., too frequent H-D substitution reactions). Conversely, observational lower D/H ratios of multiply deuterated isotopologs could stem from the negative catalytic effect of solid water. \citet{Oba2016b} experimentally verified this trend for the case of dimethyl ether and attributed it to the formation of an H-bond between the hydrogen atom of water and the oxygen atom of dimethyl ether, which strengthens the C–H bond in the dimethyl ether-water complex. However, this trend likely would have an equal effect on mono- and multiply deuterated isotopologs of O-bearing molecules. It would be interesting to experimentally test lower D/H fluxes and compare them to the observed values of I16293B to see if D/H ratios of mono-deuterated molecules will still be as high and if D/H ratios of multiply deuterated species would be lower. The presented comparison relies on the assumption that the observed gases towards I16293B are representative of freshly thermally desorbed ices (Section~\ref{spectral_analysis}), which predominantly originate from the innate prestellar core that birthed the IRAS~16293-2422 system. The experimental conditions may be more representative of the prestellar core stage with a higher availability of D. However, this still does not explain the discrepancy in experimental and observed D/H ratios of multiply deuterated species, as multiply deuterated and mono-deuterated complex organics (unlike water; \citealt{Furuya2016}) are expected to form simultaneously.

\subsection{Comparison to cometary measurements}
\label{comet}
Fig.~\ref{fig:DH_plot_exp} also includes the seven D-bearing molecules measured with the \textit{Rosetta} Orbiter Spectrometer for Ion and Neutral Analysis (ROSINA; \citealt{Balsiger2007}) instrument aboard the ESA \textit{Rosetta} spacecraft during its two-year monitoring of volatiles in the coma of comet 67P/Churyumov–Gerasimenko, hereafter 67P/C–G. The cometary D/H ratio of D$_{2}$O is higher than that of HDO, which also tentatively agrees with what is measured for IRAS~16293-2422 based on best-available estimates (Sect.~\ref{deuteration}). Within errors, it is not clear whether di-deuterated or mono-deuterated methanol has a higher D/H ratio. The main reason for this is the inability of ROSINA to distinguish CH$_{3}$OD from CH$_{2}$DOH (and CHD$_{2}$OH from CH$_{2}$DOD) by its instrumental design \citep{Drozdovskaya2021}. Fig.~\ref{fig:DH_plot_exp} shows that the cometary D/H ratio of D$_{2}$O, CHD$_{2}$OH, NH$_{2}$D, and HDS are significantly (at least one order of magnitude) lower than that in I16293B. The discrepancy for D$_{2}$O and NH$_{2}$D could be attributed to a lack of reliable protostellar measurements; however, for CHD$_{2}$OH and HDS the differences are firm and pronounced. These could be signaling differences between IRAS~16293-2422 and the core that birthed our Solar System in the availability of D atoms, which could stem from either an initially lower elemental cloud abundance of D or be set by a warmer cloud temperature. Alternatively, this may stem from chemical evolution between the Class 0 stage and the formation of comets. \citet{Paquette2021} reported the measurements of the D/H ratio in $25$ cometary dust particles that were obtained with the Cometary Secondary Ion Mass Analyser (COSIMA; \citealt{Kissel2007}) instrument aboard \textit{Rosetta}. The average value of $1.57\pm0.54\times10^{-3}$ is in range of the D/H ratios determined in cometary volatile HDO, D$_{2}$O, NH$_{2}$D, and HDS (especially if the full range of D/H ratios measured in individual particles is considered, i.e., the shaded yellow range), but not with deuterated methanol. The measured D/H ratio of the carbonaceous refractory material is more than one order of magnitude lower than that of all volatiles of IRAS~16293-2422, except the best-available estimate in HDO. Unfortunately, there are no possibilities of directly probing the composition of dust in IRAS~16293-2422. Of course, it is also relevant to realize that 67P/C-G is merely a single comet of trillions in our Solar System, which may be tracing only a specific small sector of the proto-Solar disk.


\section{Conclusions}
\label{conclusions}

This paper presents a new precise spectroscopic dataset for di-deuterated methanol, CHD$_{2}$OH, based on the global line position analysis of \citet{Coudert2021} together with an observational search towards the low-mass protostellar system IRAS~16293-2422 in the data from the Atacama Large Millimeter/submillimeter Array (ALMA) Protostellar Interferometric Line Survey (PILS; \citealt{Jorgensen2016}). Until now, CHD$_{2}$OH was the missing link in the chain of methanol deuteration. This work investigates deuteration chains of several molecules (water, formaldehyde, methanol, methyl formate, dimethyl ether, methyl cyanide) alongside one another. The main conclusions are:

   \begin{enumerate}
	    \item The best-fitting LTE parameters of CHD$_{2}$OH are $N=(1.78\pm0.36)\times10^{17}$~cm$^{-2}$ and $T_{\text{ex}}=90.0\pm9$~K towards a one-beam offset position from IRAS~16293-2422~B in the SW direction ($0.5\arcsec$ or $70$~au away from the dust peak of B), which yields a D/H ratio $7.7\pm1.2\%$ with statistical correction. Towards a (slightly more than) one-beam offset position from IRAS~16293-2422~A in the NE direction ($0.6\arcsec$ or $85$~au away from the dust peak of A), the best-fitting LTE parameters of CHD$_{2}$OH are $N=(2.22\pm0.44)\times10^{17}$~cm$^{-2}$ and $T_{\text{ex}}=127\pm13$~K, which yields a D/H ratio $7.5\pm1.1\%$ with statistical correction.
      \item In IRAS~16293-2422~B, the hot corino D/H ratios of all available molecules lie in the $0.55-16\%$ range. This includes statistical corrections and is based on $36$ data points (which include mono-, di-, and tri-deuterated molecules).
      \item No trends in the D/H ratios with functional groups are seen, nor with families (i.e., CHO-, N-, or S-bearing) when considering all D-bearing molecules. When considering solely mono-deuterated species, D/H ratios tentatively seem to be lower for N-bearing molecules and O-bearing species with $<6$ atoms in comparison to S-bearing molecules and O-bearing species with $>6$ atoms. This has already been observed in past papers for the O-bearing species and may be related to the differences in formation timescales.
      \item The D/H ratios of di-deuterated molecules relative to their mono-deuterated counterparts are consistently and significantly higher than the D/H ratios of mono-deuterated isotopologs relative to the non-deuterated species. Moreover, these fractionation levels of mono-deuterated formaldehyde, methanol, methyl formate, dimethyl ether, and methyl cyanide are remarkably similar ($10-33\%$ range with a mean value of $20.7\pm6.2\%$) and agree within error bars. This trend is also starting to be seen in other low-mass star-forming systems as inventories of mono- and di-deuterated molecules in other sources are expanding.
			\item Tentatively, the best-available estimates of water deuteration suggest D$_{2}$O to have a D/H ratio relative to HDO that is a factor of $\sim2$ lower than of other di-deuterated molecules relative to their respective mono-deuterated counterparts. The D/H ratio of HDO relative to H$_{2}$O is also lower than of all other molecules considered in low-mass star-forming regions. This is also seen in other sources, not just IRAS~16293-2422.
			\item When comparing D/H ratios of detected D-bearing molecules in IRAS~16293-2422~A and~B with those in other low-mass Class~0 and I sources (L483, SVS13-A, NGC~1333-IRAS2A, -IRAS4A, and -IRAS4B), it does not appear that IRAS~16292-2422 is in any way abnormally deuterium-rich.
			\item Comparing D/H ratios of deuteration chains of formaldehyde and methanol in IRAS~16293-2422~B with those obtained in laboratory experiments \citep{Nagaoka2005, Nagaoka2007, Hidaka2009} shows that higher levels of fractionation in multiply deuterated isotopologs are obtained at a constant D/H input flux, which implies that this is not an indicator of multiply deuterated isotopologs forming in more D-rich conditions. Rather, this is a natural consequence of successive H-D substitution reactions (i.e., H-abstractions followed by D additions).
			\item Comparing D/H rations of molecules in IRAS~16293-2422~B with those measured in volatiles of comet 67P/Churyumov–Gerasimenko with \textit{Rosetta}-ROSINA shows a firm and pronounced discrepancy in di-deuterated methanol and HDS (and, tentatively, in D$_{2}$O and NH$_{2}$D). The disagreement is less pronounced in mono-deuterated methanol due to larger cometary error bars and, tentatively, in HDO. The D/H ratio of the carbonaceous refractory cometary material measured with \textit{Rosetta}-COSIMA is more than one order of magnitude lower than that of all volatiles of IRAS~16293-2422~B except for HDO (tentatively). The Solar System's natal cloud may have had a lower initial elemental abundance of D or may have been warmer than the cloud of IRAS~16293-2422; or the cloud material may have been partially reprocessed prior to incorporation into a comet.
   \end{enumerate}

The analysis presented in this paper highlights the importance of accurate molecular spectroscopy for studies of isotopic fractionation. Concurrently, such studies have the potential to shed new light on the formation of organic molecules and their role in the chemistry of the early Solar System. Future studies of successive deuteration in large samples of protostellar sources, including high-mass star-forming regions, will test its role as a physicochemical tracer under a wider range of physical conditions.


\begin{acknowledgements}
This work is supported by the Swiss National Science Foundation (SNSF) Ambizione grant no. 180079, the Center for Space and Habitability (CSH) Fellowship, and the IAU Gruber Foundation Fellowship. AC is supported by the European Research Council (ERC) under the European Union's Horizon 2020 research and innovation programme through ERC Starting Grant ``Chemtrip'' (grant agreement \textnumero~949278). J.K.J. is supported by the Independent Research Fund Denmark (grant number 0135-00123B).
\\
\\
The authors would like to thank Daniel Harsono and Florian Reinhard for useful discussions about the MCMC modeling routines that have been developed for this work. This work benefited from discussions held with the international team \#461 ``Provenances of our Solar System's Relics'' (team leaders Maria N. Drozdovskaya and Cyrielle Opitom) at the International Space Science Institute, Bern, Switzerland.
\end{acknowledgements}

\bibliographystyle{aa} 
\bibliography{CHD2OH_IRAS16293_bib} 


\begin{appendix} 
\section{Additional figures and detailed assumptions}
\label{additional_figures}

\subsection{Main isotopolog column densities in I16293}
For ketene, formic acid, formamide, isocyanic acid, and cyanamide, the column densities of the main isotopologs are based on their minor carbon isotopologs and an assumed local ISM $^{12}$C/$^{13}$C ratio \citep{Jorgensen2018, Coutens2016, Ligterink2017, Coutens2018a}. For methyl cyanide, the column density was derived based in the vibrationally excited $v_{8}=1$ state \citep{Calcutt2018a}. For methanol, the column density is based on its minor oxygen isotopolog and an assumed local ISM $^{16}$O/$^{18}$O ratio \citep{Jorgensen2016}. For hydrogen sulfide, the column density is merely an estimate based on the jointly detected D-bearing isotopologs and past single dish data (section~$3.6$ of \citealt{Drozdovskaya2018}; however, on-going analysis of additional observations is constraining $N(\text{H}_{2}\text{S})$ more accurately and showing it to be very close to the estimated value used here; Kushwahaa et al. in prep.). For formaldehyde, ethanol, dimethyl ether, methyl formate, glycolaldehyde, acetaldehyde, and thioformaldehyde, the column densities of the main isotopologs are more reliable, because they were determined based on optically thin lines alone of the main isotopologs themselves \citep{Persson2018, Jorgensen2018, Manigand2019, Drozdovskaya2018}.

\subsection{L483}
\citet{Agundez2019} carried out a line survey in the $80-116$~GHz frequency range with IRAM~30~m towards the cold dense protostellar core L483 harboring a Class 0 source. The cold nature of the source reduces the number of lines that are observable, e.g., just one line of CHD$_{2}$OH is detected (table~$1$ of \citealt{Agundez2019}), which in turn makes it difficult to constrain $T_{\text{ex}}$ and $N$. Nevertheless, several di-deuterated species have been detected, including di-deuterated formaldehyde and methanol that are in common with I16293B (\citealt{Persson2018} and this work)\footnote{Likely, di-deuterated cyclopropenylidene (c-C$_{3}$D$_{2}$) is common to both sources as well, although the detection is yet to be confirmed following the labeling in \citealt{MartinDomenech2016a}. Di-deuterated thioformaldehyde and propyne (methyl acetylene) have not been detected in the hot corino of I16293B at the sensitivity of the ALMA-PILS data \citep{Drozdovskaya2018, Calcutt2019}, but are detected in L483 \citep{Agundez2019, Agundez2021b}.}. \citet{Jacobsen2019, Jensen2019}, and \citet{Jensen2021} targeted the embedded protostar of this system with ALMA and detected HDO, D$_{2}$O, CH$_{3}$OD, and CH$_{2}$DOH in the hot inner regions. Fig.~\ref{fig:DH_plot_protostars} shows that the ratio of CH$_{2}$DOH/CH$_{3}$OH in the hot corino is in close agreement with this ratio in the envelope\footnote{Note that \citet{Jacobsen2019} determines $N$(CH$_{3}$OH) based on optically thin lines of methanol itself, which are accessible in this source at the $\sim350$~GHz frequencies targeted in that work. Meanwhile, \citet{Agundez2019} derived $N$(CH$_{3}$OH) based on $N$($^{13}$CH$_{3}$OH) and the local ISM $^{12}$C/$^{13}$C~$=68$ ratio, because the methanol lines at $\sim100$~GHz are optically thick. The assumed value of $68$ has been supported by the column density of $^{13}$CH$_{3}$OH derived in \citet{Jacobsen2019}. Note that the same approach was also undertaken by \citet{Agundez2019} to derive $N$(H$_{2}$CO) based on $N$(H$_{2}^{13}$CO). $N$(NH$_{3}$) is the best-available value from \citet{Anglada1997}. \citet{Jensen2019} and \citet{Jensen2021} derived $N$(H$_{2}$O) based on $N$(H$_{2}^{18}$OH) and the local ISM $^{16}$O/$^{18}$O~$=560$ ratio. In Fig.~\ref{fig:DH_plot_protostars}, the error bars on the values from \citet{Jacobsen2019, Jensen2019, Jensen2021, Agundez2019, Agundez2021b} correspond to uncertainties of $30\%$.}, suggesting that the envelope abundance ratios of \citet{Agundez2019} may be representative of the hot corino abundance ratios for other D-bearing molecules in L483.

\subsection{SVS13-A}
\citet{Bianchi2017a, Bianchi2019b} observed the Class I SVS13-A protostar with IRAM~30~m and detected CHD$_{2}$OH at two unique frequencies around $208$~GHz. Simultaneously, deuterated formaldehyde (HDCO, D$_{2}$CO), mono-deuterated methanol (CH$_{2}$DOH, CH$_{3}$OD), and mono-deuterated thioformaldehyde (HDCS) were also detected. Consequently, all D-bearing molecules of SVS13-A are common to I16293B. Methanol observations of SVS13-A were interpreted with two components: a compact hot corino and an extended colder envelope, both of which are plotted in Fig.~\ref{fig:DH_plot_protostars}. For the data points of SVS13-A, the D/H ratios and errors bars of HDCO, D$_{2}$CO, and CH$_{2}$DOH are taken directly from table~$3$ of \citet{Bianchi2017a}, which are based on column densities calculated with the $T_{\text{rot}}$ of the respective $^{13}$C-isotopologs. The D/H ratios stemming from CH$_{3}$OD and CHD$_{2}$OH have been calculated based on the provided estimated column densities of these two species in \citet{Bianchi2017a}, and $N$(CH$_{3}$OH) as calculated from $N$($^{13}$CH$_{3}$OH) and $^{12}$C/$^{13}$C~$=68$ (\citealt{Milam2005}; \citealt{Bianchi2017a} quote $^{12}$C/$^{13}$C~$=86$, which is likely a typo). The error bars correspond to uncertainties of $73\%$, as estimated for the worst case scenario from the uncertainties tabulated in table~$3$ of \citet{Bianchi2017a} for D/H ratios from other molecules. The D/H ratio and errors bars of HDCS are taken directly from figure~$4$ of \citet{Bianchi2019b}.

\subsection{NGC~1333-IRAS2, -IRAS4A, and -IRAS4B}
\citet{Parise2006} detected CHD$_{2}$OH, CH$_{2}$DOH, and CH$_{3}$OD towards NGC~1333-IRAS2, -IRAS4A, and -IRAS4B with IRAM~30~m observations. The former two have subsequently been observed with the IRAM Plateau de Bure Interferometer (IRAM-PdBI) by \citet{Taquet2015, Taquet2019}, which allowed the column densities to be derived on smaller spatial scales of the hot corinos. The methanol and methyl cyanide column densities shown in Fig.~\ref{fig:DH_plot_protostars} for IRAS2A and IRAS4A stem from these PdBI observations, which are associated with the analysis of optically thin CH$_{3}$OH lines (as verified by $^{13}$CH$_{3}$OH line analysis and the derivation of the standard local ISM $^{12}$C/$^{13}$C~$=70$ ratio). For IRAS4B, only single dish estimates of the column densities of deuterated methanol are available from \citet{Parise2006}. The column density of HDO in IRAS2A, IRAS4A, and IRAS4B stems from PdBI observations of \citet{Persson2014} and of H$_{2}$O from PdBI observations by \citet{Persson2012} of H$_{2}^{18}$O upon assumption of the standard $^{16}$O/$^{18}$O~$=560$ ratio. These values are in agreement with those derived on larger spatial scales from \textit{Herschel} observations of IRAS4A and IRAS4B (\citealt{Coutens2013b}; not shown). D$_{2}$O has been detected by \citet{Coutens2014a} in IRAS2A by PdBI observations. The HDO and H$_{2}^{18}$O column densities of \citet{Coutens2014a} are within a factor of $2$ of those derived in \citet{Persson2012, Persson2014}. D/H ratios of HDCO and D$_{2}$CO (and the associated H$_{2}$CO abundance) stem from the JCMT Spectral Legacy Survey (SLS; \citealt{Koumpia2016, Koumpia2017}). The column densities of NH$_{2}$D and NH$_{3}$ stem from \citet{WoottenJeffrey1993} and \citet{Blake1995}, respectively, as presented in \citet{ShahWootten2001}, which is a combination of VLA, CSO, and JCMT data. The error bars correspond to $30\%$ uncertainties, based on the errors reported in the respective publications.

\subsection{Binarity}
All NGC~1333 sources appear to be part of binaries based on the latest ALMA and VLA observations. IRAS4A is a binary source of A1 and A2 with a separation of $1.83\arcsec$ ($\sim547$~au)  \citep{Tobin2016a, LopezSepulcre2017, DeSimone2020a}. \citet{Sahu2019} derived the column density of CH$_{2}$DOH and of CH$_{3}$OH based on $^{13}$CH$_{3}$OH, yielding D/H ratios of $2.0\times10^{-3}$ and $1.6\times10^{-2}$ for A2 and A1, respectively. These values are not far off from those of \citet{Taquet2019} for the binary as a whole (not shown in Fig.~\ref{fig:DH_plot_protostars}). IRAS2A is a binary source of VLA1 and VLA2 with a separation of $0.62\arcsec$ ($\sim185$~au) \citep{Maret2014, Codella2014a, Tobin2015a, Tobin2016a}. IRAS4B (sometimes labeled BI) has a binary component B' (or BII) that is 10.654\arcsec ($\sim3~186$~au) away \citep{Sakai2012, Anderl2016, Tobin2016a}. Here, a distance to NGC1333 of $299$~pc is adopted \citep{Zucker2018}. The data points in Fig.~\ref{fig:DH_plot_protostars} do not separate A1 from A2 in IRAS4A, nor VLA1 from VLA2 in IRAS2A, but IRAS4B data do not include contributions from IRAS4B'. All these sources are classified as Class 0s. L483 is an isolated source at a distance of $\sim200$~pc (\citealt{DameThaddeus1985}; although there is quite some uncertainty here as discussed in \citealt{Jacobsen2019}). SVS13-A is binary source of VLA4A and VLA4B with a separation of $0.30\arcsec$ ($\sim90$~au), and a third source VLA3 (or SVS13A2) that is $5.314\arcsec$ ($\sim1~589$~au) away \citep{Tobin2016a, Codella2021}. The data in Fig.~\ref{fig:DH_plot_protostars} for SVS13-A are a combination of VLA4A and VLA4B, and may even contain contributions from VLA3. SVS13-A is also located in NGC1333.

\begin{figure}
	\centering
	\includegraphics[width=\hsize]{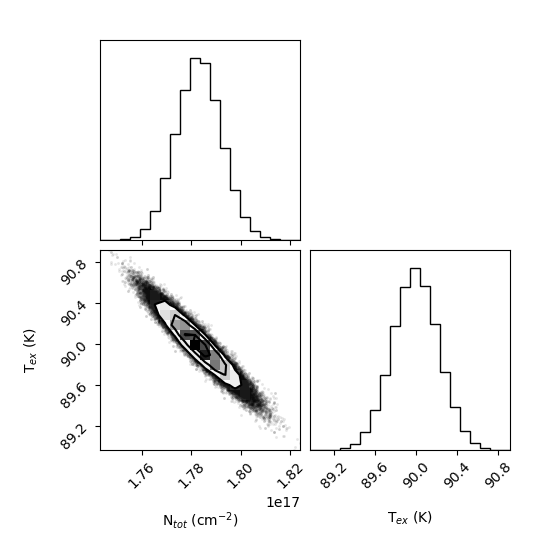}
	\caption{Corner plot of the second MCMC run of the CHD$_{2}$OH computation for source B.}
	\label{fig:CD2HOH_v1_corner2}
\end{figure}

\begin{figure}
	\centering
	\includegraphics[width=\hsize]{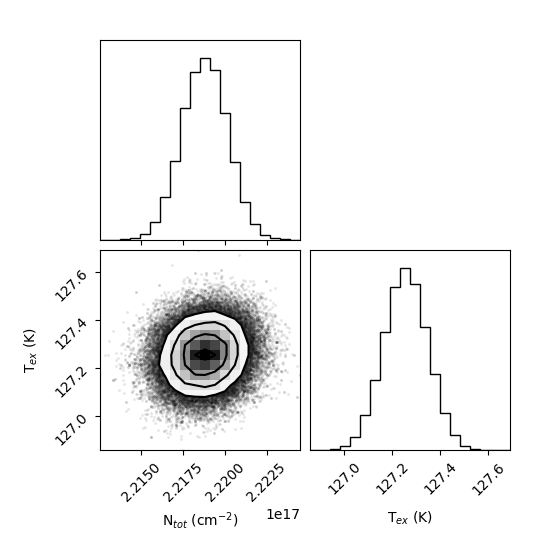}
	\caption{Corner plot of the second MCMC run of the CHD$_{2}$OH computation for source A.}
	\label{fig:CD2HOH_v1_corner2_A}
\end{figure}

\begin{figure}
	\centering
	\includegraphics[width=\hsize]{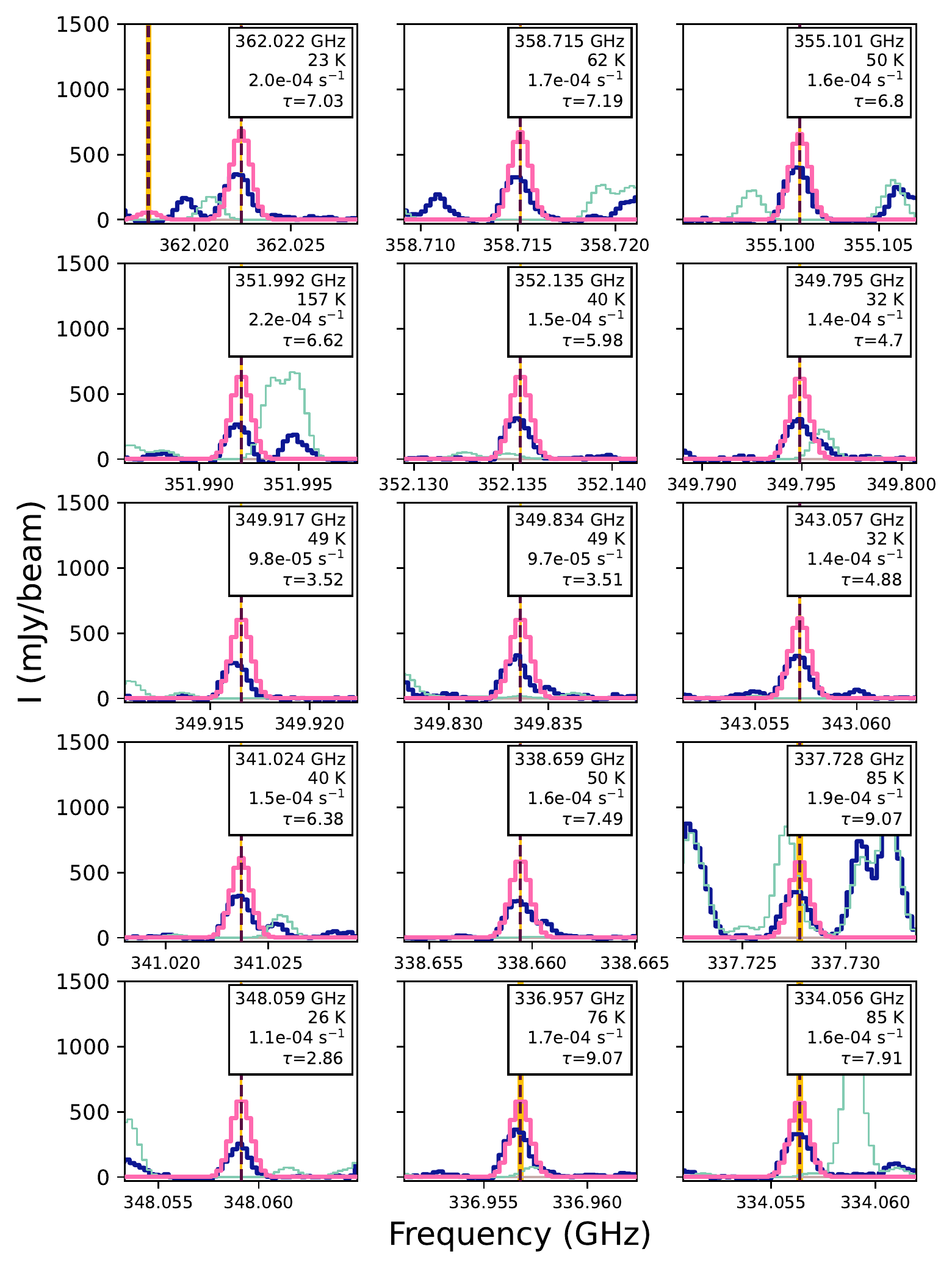}
	\caption{Another selection of some of the strongest $15$ lines of CHD$_{2}$OH observed (all optically thick). None of these have been used in the MCMC spectral fitting. The observed spectrum is in dark blue, the ``reference'' spectrum is in turquoise, and the best-fitting synthetic spectrum is in pink. The rest frequency, $E_{\text{up}}$ (K), $A_{ij}$ (s$^{-1}$), and optical depth (for the best-fitting parameters) are shown in the right corner of each panel. The rest frequency is indicated with a vertical dashed line, and the filled yellow region corresponds to the uncertainty on that line frequency.}
	\label{fig:CD2HOH_panel_plot_top15}
\end{figure}

\begin{figure}
	\centering
	\includegraphics[width=\hsize]{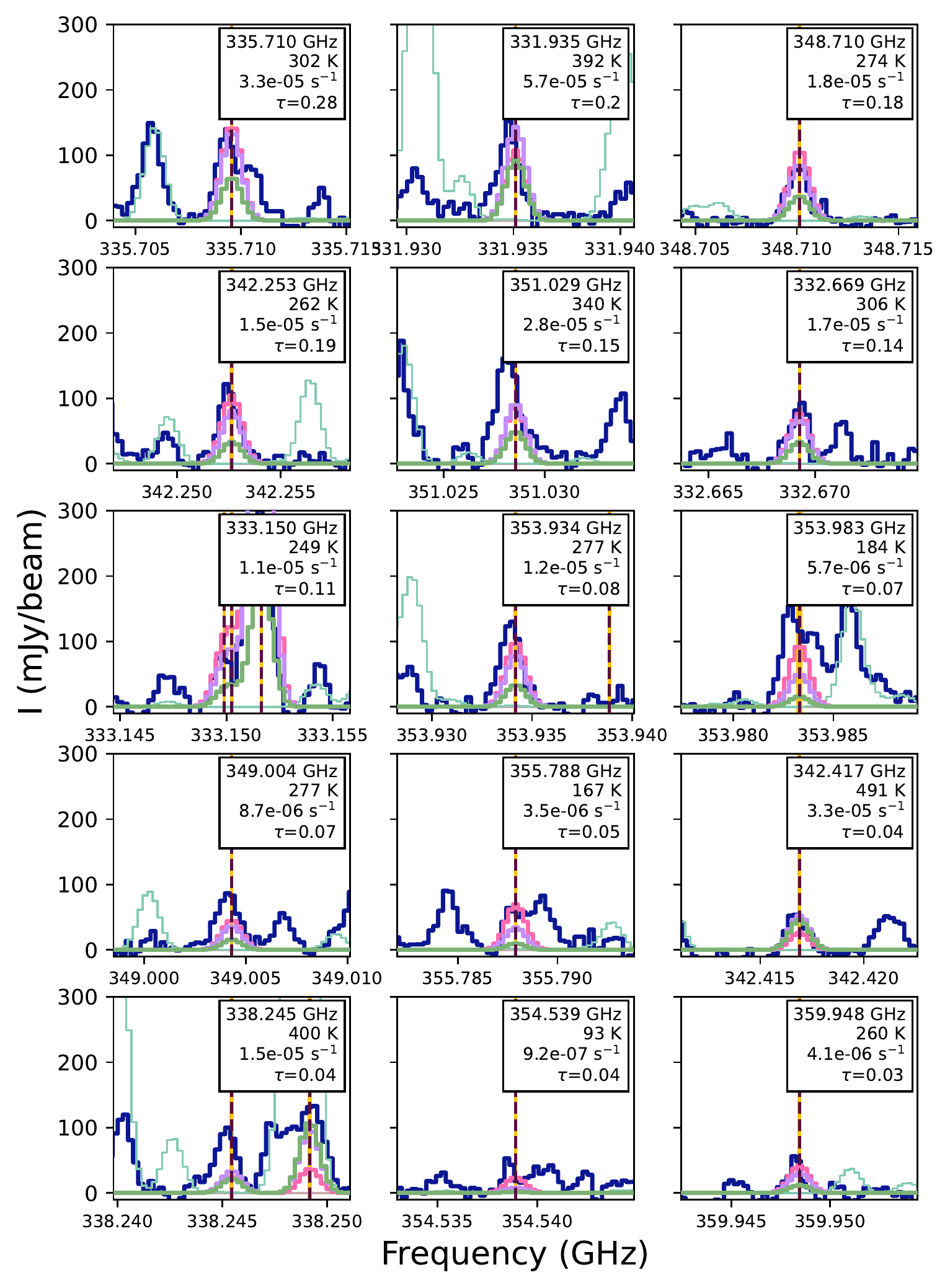}
	\caption{Same selection of some of the strongest $15$ lines of CHD$_{2}$OH out of the $105$ used for synthetic spectral fitting as in Fig.~\ref{fig:CD2HOH_panel_plot}, but now with also the best-fitting models for fixed $T_{\text{ex}}=150$ and $300$~K overlaid in lilac and green, respectively. The observed spectrum is in dark blue, the ``reference'' spectrum is in turquoise, and the best-fitting synthetic spectrum ($T_{\text{ex}}=90$~K) is in pink. The rest frequency, $E_{\text{up}}$ (K), $A_{ij}$ (s$^{-1}$), and optical depth (for the best-fitting parameters) are shown in the right corner of each panel. The rest frequency is indicated with a vertical dashed line, and the filled yellow region corresponds to the uncertainty on that line frequency. Note that the lines at $353.934$, $353.983$, $355.788$, and $359.948$~GHz are in fact overlapping double transitions.}
	\label{fig:CD2HOH_panel_plot_150K_300K}
\end{figure}

\begin{figure}
	\centering
	\includegraphics[width=\hsize]{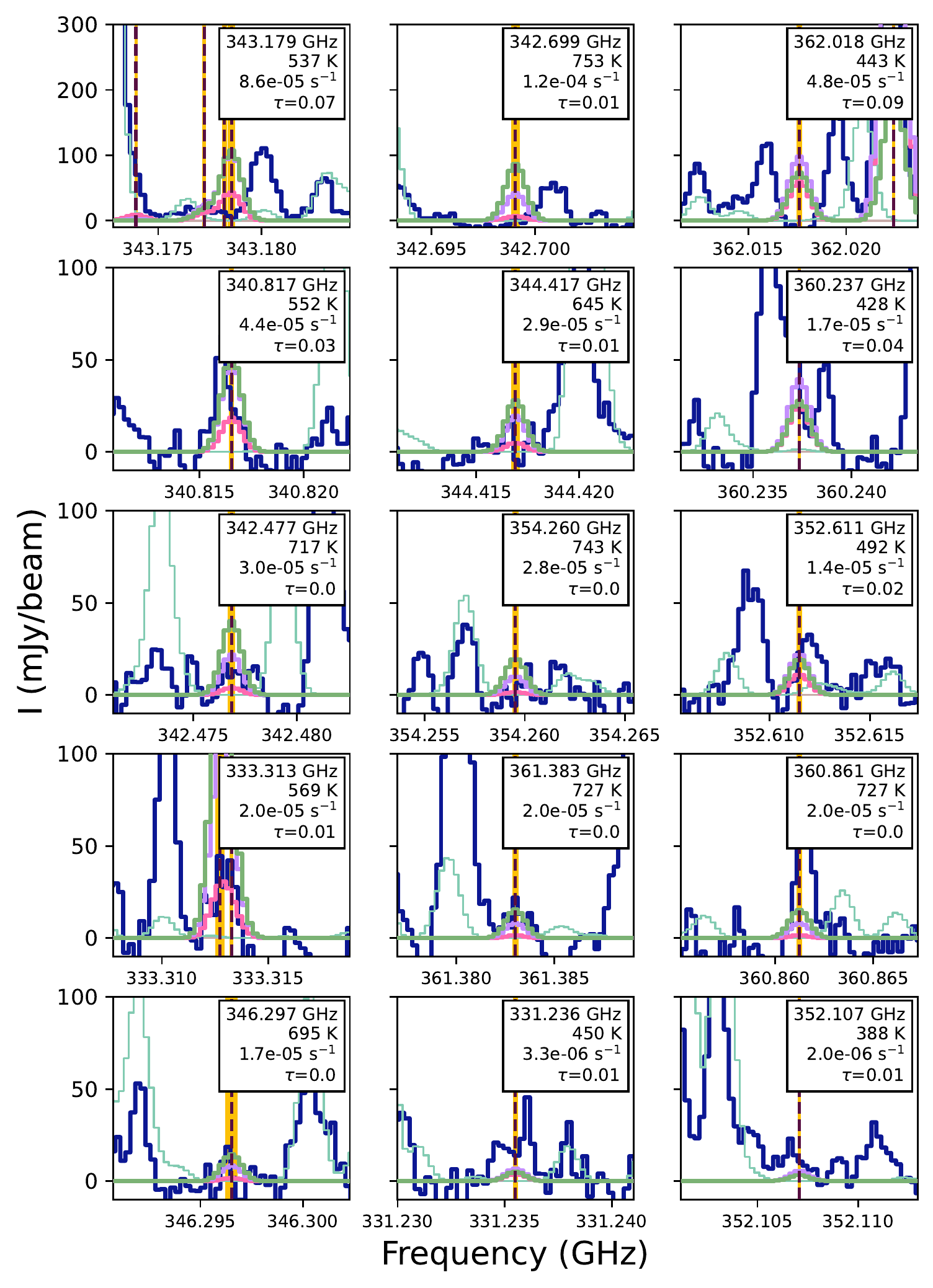}
	\caption{Highly excited lines of CHD$_{2}$OH (out of the $105$ used for synthetic spectral fitting) that drive the best-fitting excitation temperature to lower values (Sect.~\ref{line_profiles}). The observed spectrum is in dark blue, the ``reference'' spectrum is in turquoise, and the best-fitting synthetic spectrum ($T_{\text{ex}}=90$~K) is in pink. The best-fitting models for fixed $T_{\text{ex}}=150$ and $300$~K are overlaid in lilac and green, respectively. The rest frequency, $E_{\text{up}}$ (K), $A_{ij}$ (s$^{-1}$), and optical depth (for the best-fitting parameters) are shown in the right corner of each panel. The rest frequency is indicated with a vertical dashed line, and the filled yellow region corresponds to the uncertainty on that line frequency. Note that the lines at $342.477$ and $333.313$~GHz are in fact overlapping double transitions.}
	\label{fig:CD2HOH_panel_plot_150K_300K_non_det}
\end{figure}

\begin{figure}
	\centering
	\includegraphics[width=\hsize]{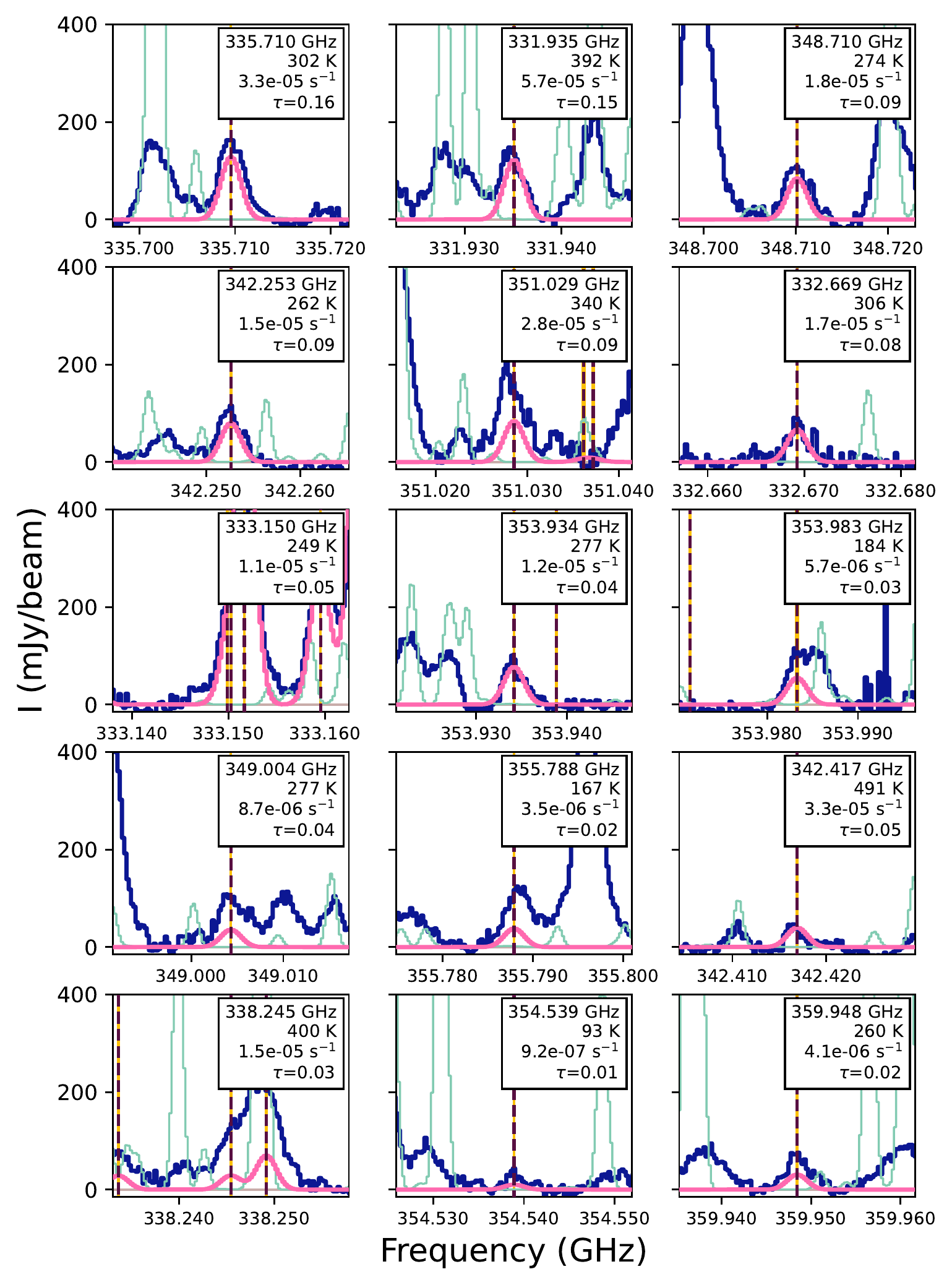}
	\caption{Same selection of some of the strongest $15$ lines of CHD$_{2}$OH out of the $105$ used for synthetic spectral fitting as in Fig.~\ref{fig:CD2HOH_panel_plot}, but now for source A. The observed spectrum is in dark blue, the best-fitting synthetic spectrum of is in pink. The ``reference'' spectrum is in turquoise, but note that the reference spectrum belongs to source B (shown to check for blending and to facilitate a comparison with Fig.~\ref{fig:CD2HOH_panel_plot}). The rest frequency, $E_{\text{up}}$ (K), $A_{ij}$ (s$^{-1}$), and optical depth (for the best-fitting parameters) are shown in the right corner of each panel. The rest frequency is indicated with a vertical dashed line, and the filled yellow region corresponds to the uncertainty on that line frequency. Note that the lines at $353.934$, $353.983$, $355.788$, and $359.948$~GHz are in fact overlapping double transitions.}
	\label{fig:CD2HOH_panel_plot_A}
\end{figure}

\begin{figure*}
	\centering
	\includegraphics[width=\hsize]{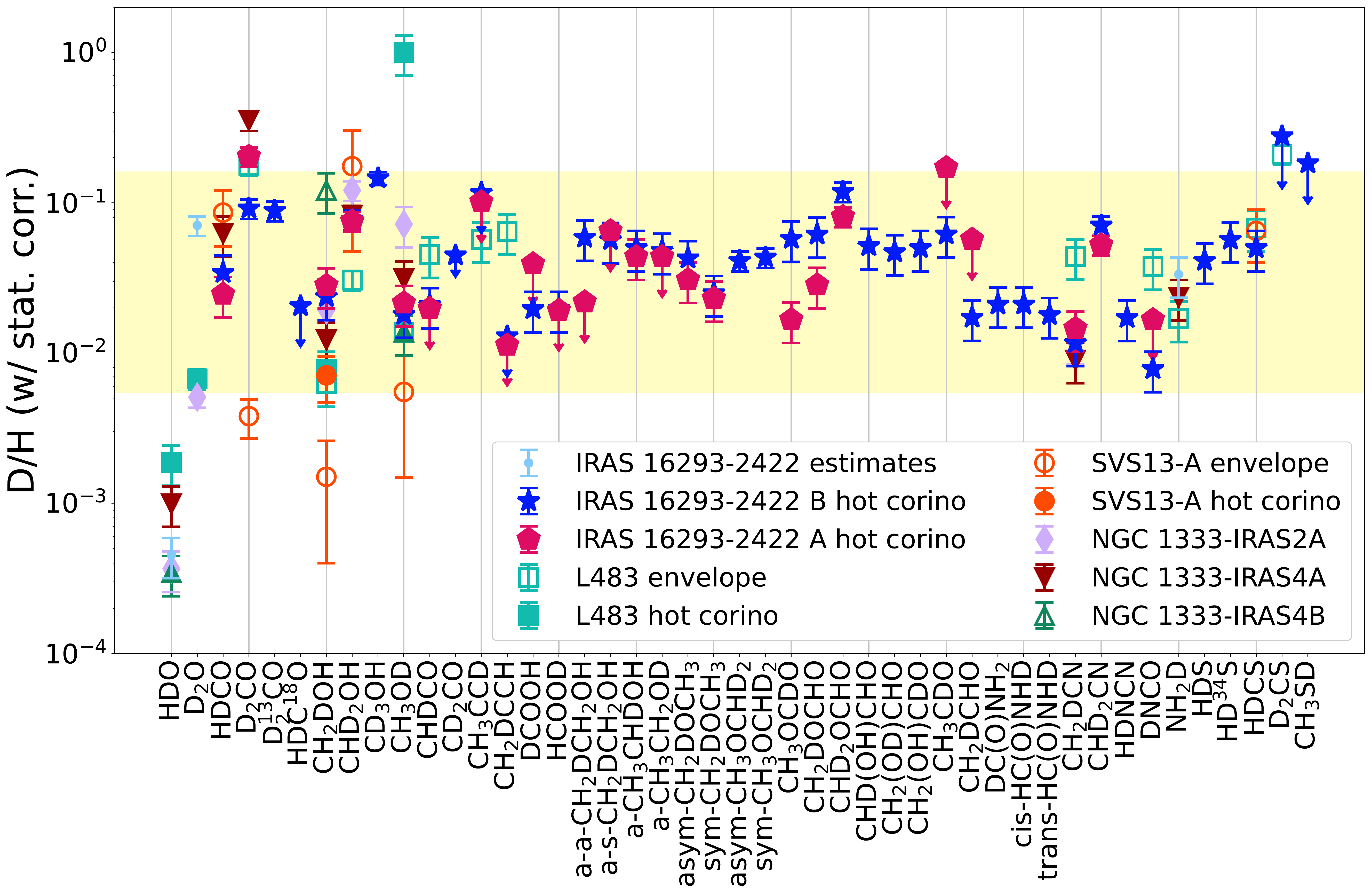}
	\caption{D/H ratio as measured in all the D-bearing molecules detected in the hot corinos of IRAS~16293-2422~A and~B, including the respective statistical corrections (Appendix~\ref{statistical_corrections}). Available D/H ratios measured in other low-mass star-forming systems are also shown, see the main text for details.}
	\label{fig:DH_plot_protostars_A}
\end{figure*}

\begin{figure*}
	\centering
	\includegraphics[width=\hsize]{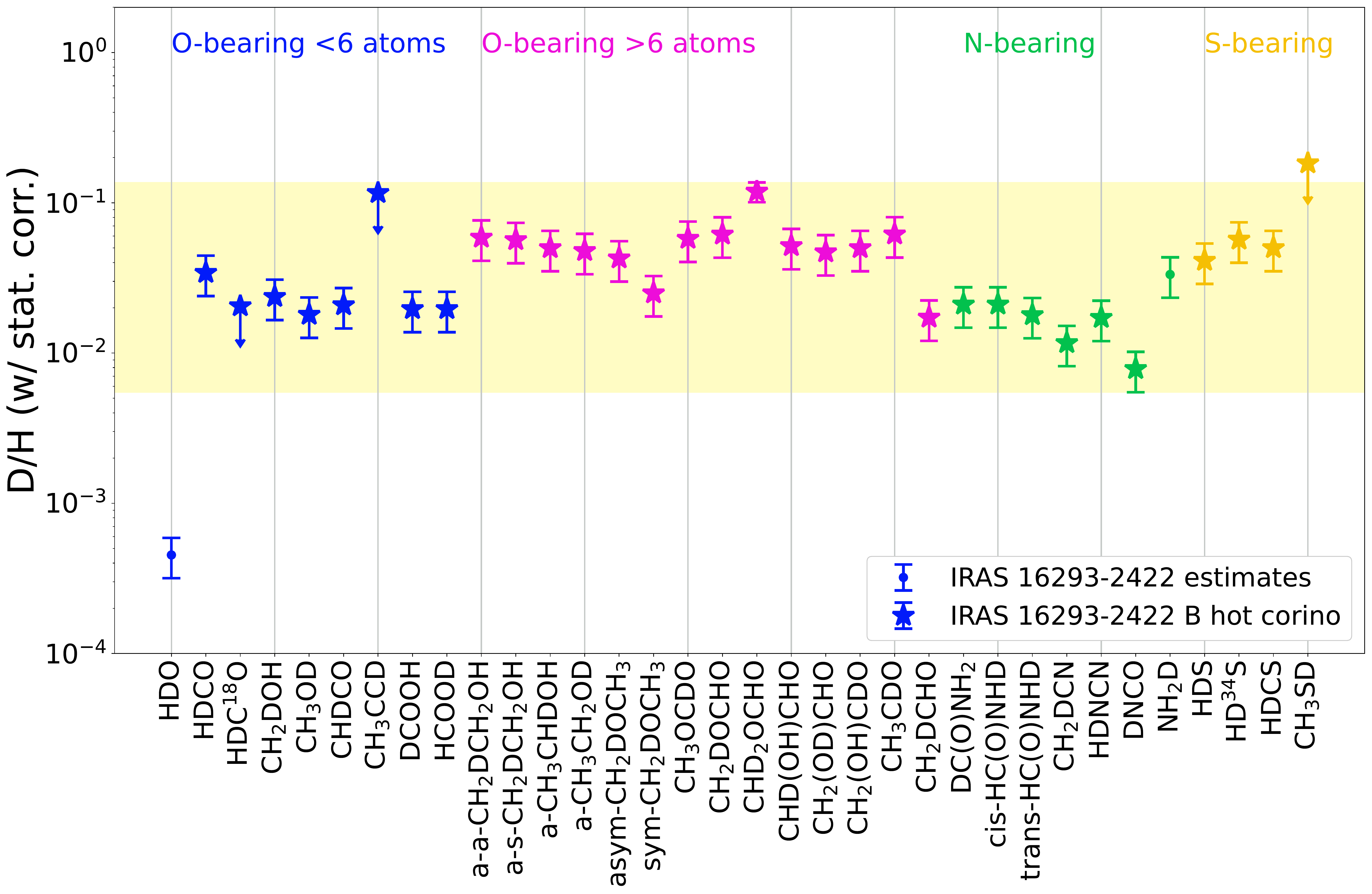}
	\caption{D/H ratio as measured in all the mono-deuterated molecules detected in the hot corino of IRAS~16293-2422~B, including the respective statistical corrections (Appendix~\ref{statistical_corrections}). This is a subset of the data points shown in Fig.~\ref{fig:DH_plot_protostars}. The data points are blue if the molecules are O-bearing and have no more than 6 atoms, fuchsia if the molecules are O-bearing and contain more than 6 atoms, green if the molecules are N-bearing, and gold if the molecules are S-bearing.}
	\label{fig:DH_plot_protostars_mono}
\end{figure*}

\begin{figure}
	\centering
	\includegraphics[width=\hsize]{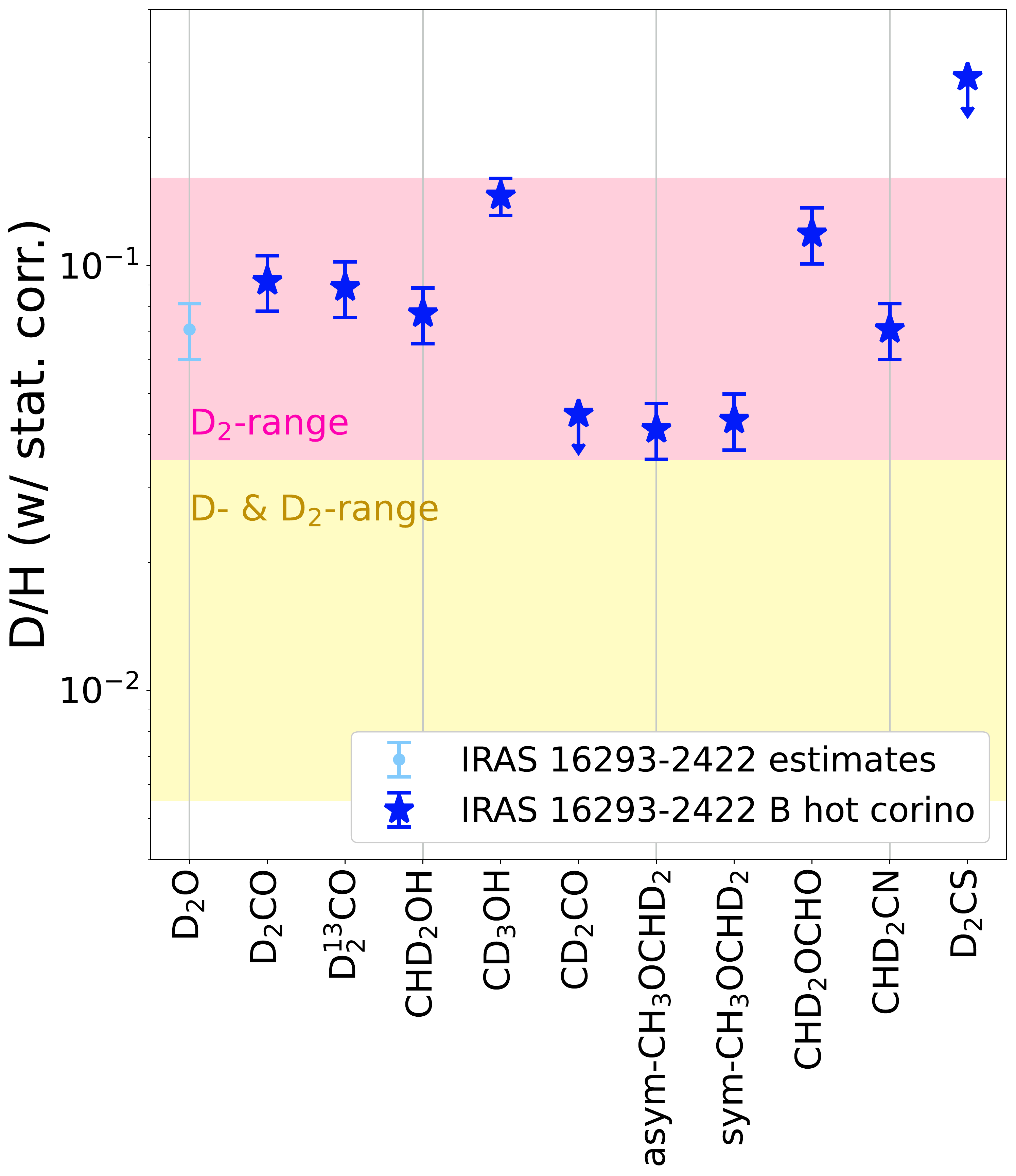}
	\caption{D/H ratio as measured in all the di-deuterated molecules detected in the hot corino of IRAS~16293-2422~B, including the respective statistical corrections (Appendix~\ref{statistical_corrections}). This is a subset of the data points shown in Fig.~\ref{fig:DH_plot_protostars}.}
	\label{fig:DH_plot_protostars_di}
\end{figure}

\clearpage

\section{Integrated intensity maps details}
\label{integrated_intensity_maps_details}
\begin{figure*}
	\centering
	\includegraphics[width=0.9\hsize]{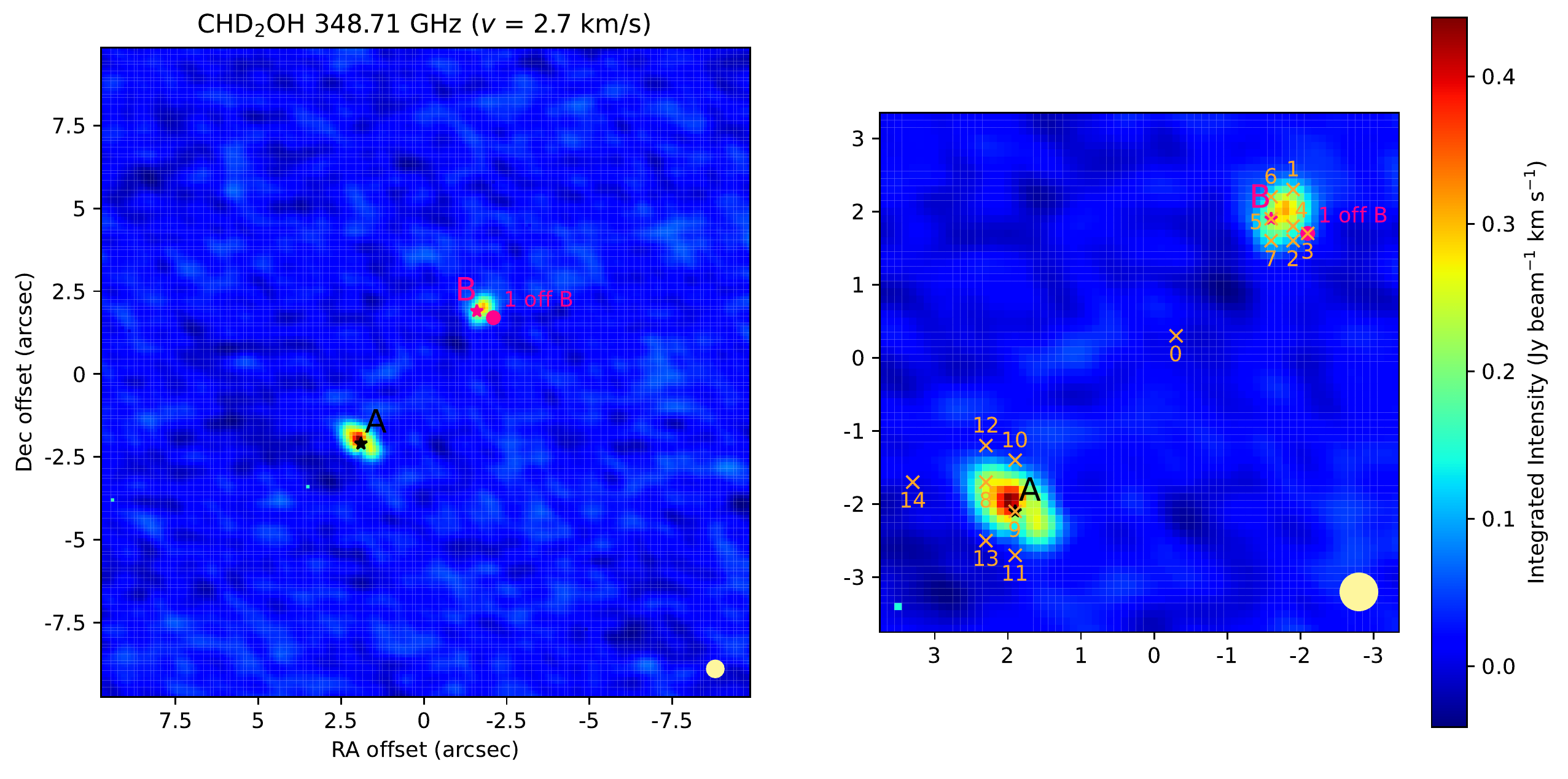}
	\caption{Integrated intensity maps of CHD$_{2}$OH (moment 0 maps) in classical (non-VINE) method. The origin of the axes corresponds to RA, Dec (J2000) of 16h32m22.72s, -24d28$\arcmin$34.3$\arcsec$. The $0.5\arcsec$ beam size is indicated in the bottom right by a yellow circle.}
	\label{fig:CD2HOH_348.7101234_mom0_2.7_multi}
\end{figure*}

\begin{figure*}
	\centering
	\includegraphics[width=0.9\hsize]{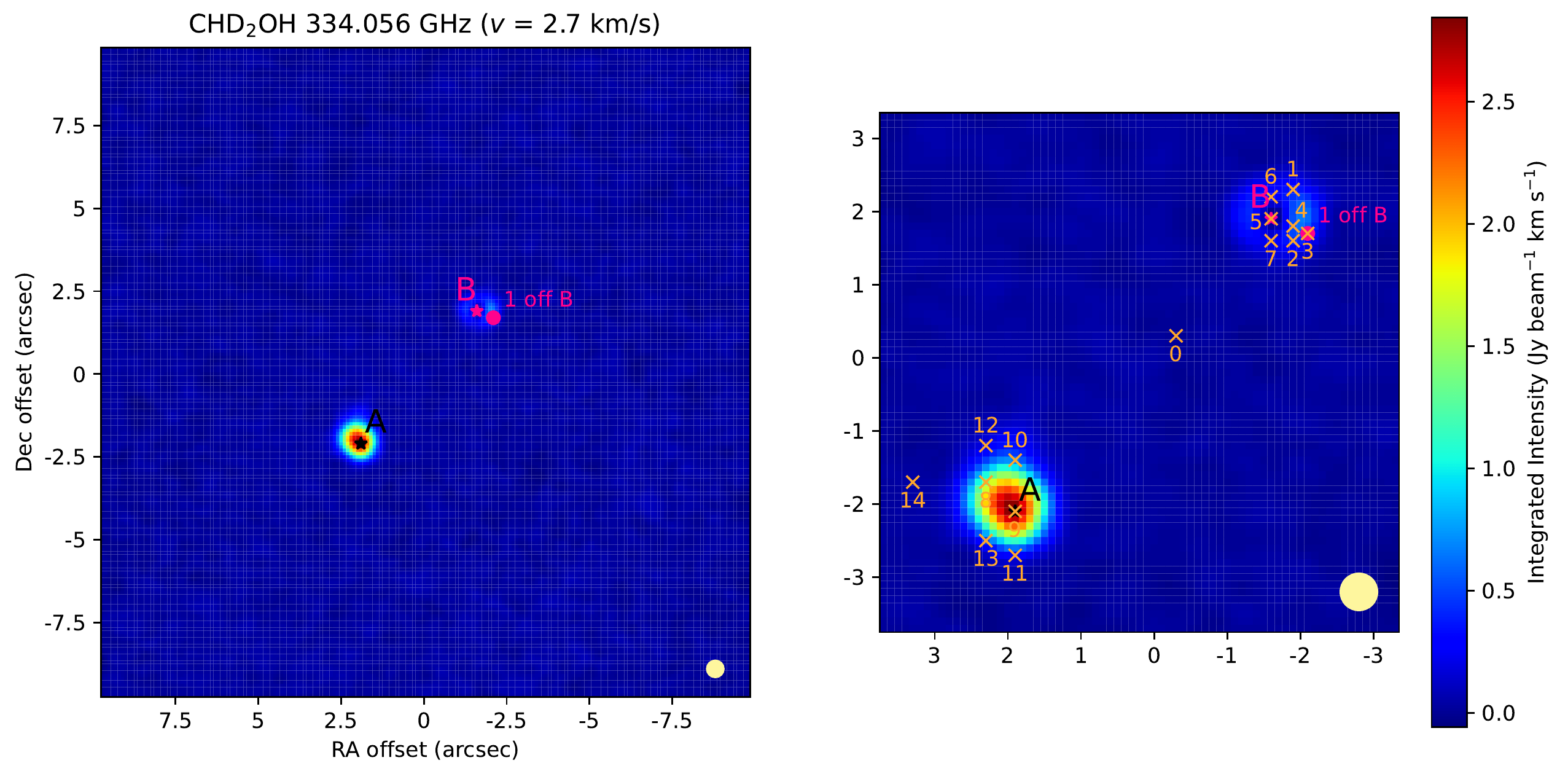}
	\caption{Integrated intensity maps of CHD$_{2}$OH (moment 0 maps) in classical (non-VINE) method. The origin of the axes corresponds to RA, Dec (J2000) of 16h32m22.72s, -24d28$\arcmin$34.3$\arcsec$. The $0.5\arcsec$ beam size is indicated in the bottom right by a yellow circle.}
	\label{fig:CD2HOH_334.056369_mom0_2.7_multi}
\end{figure*}

In this Appendix, some additional details in connection to integrated intensity maps (moment 0 maps) are presented. Fig.~\ref{fig:CD2HOH_348.7101234_mom0_2.7_multi} displays the ``classical'' (non-VINE) integrated intensity map, in which a constant $v_{\text{LSR}}$ of $2.7$~km~s$^{-1}$ is applied. The left panel shows the full field of view (FOV) of the observations and the right panel shows a zoom-in of the two sources. This panel can be compared to the VINE map of the same line in Fig.~\ref{fig:CD2HOH_348.7101234_vine_multi_paper}. Only minor differences are seen in this case. Fig.~\ref{fig:CD2HOH_334.056369_mom0_2.7_multi} displays the ``classical'' integrated intensity map of the line with a VINE map in Fig.~\ref{fig:CD2HOH_334.056369_vine_multi_paper}. In this case, the differences between the two for source A are more significant. The morphology of the emission is somewhat more extended in the VINE map. Also, the peak integrated intensity is higher. Both of these points indicate that the VINE map is doing a better job at recovering the line emission near source A.

\begin{figure*}
	\centering
	\includegraphics[height=0.45\vsize]{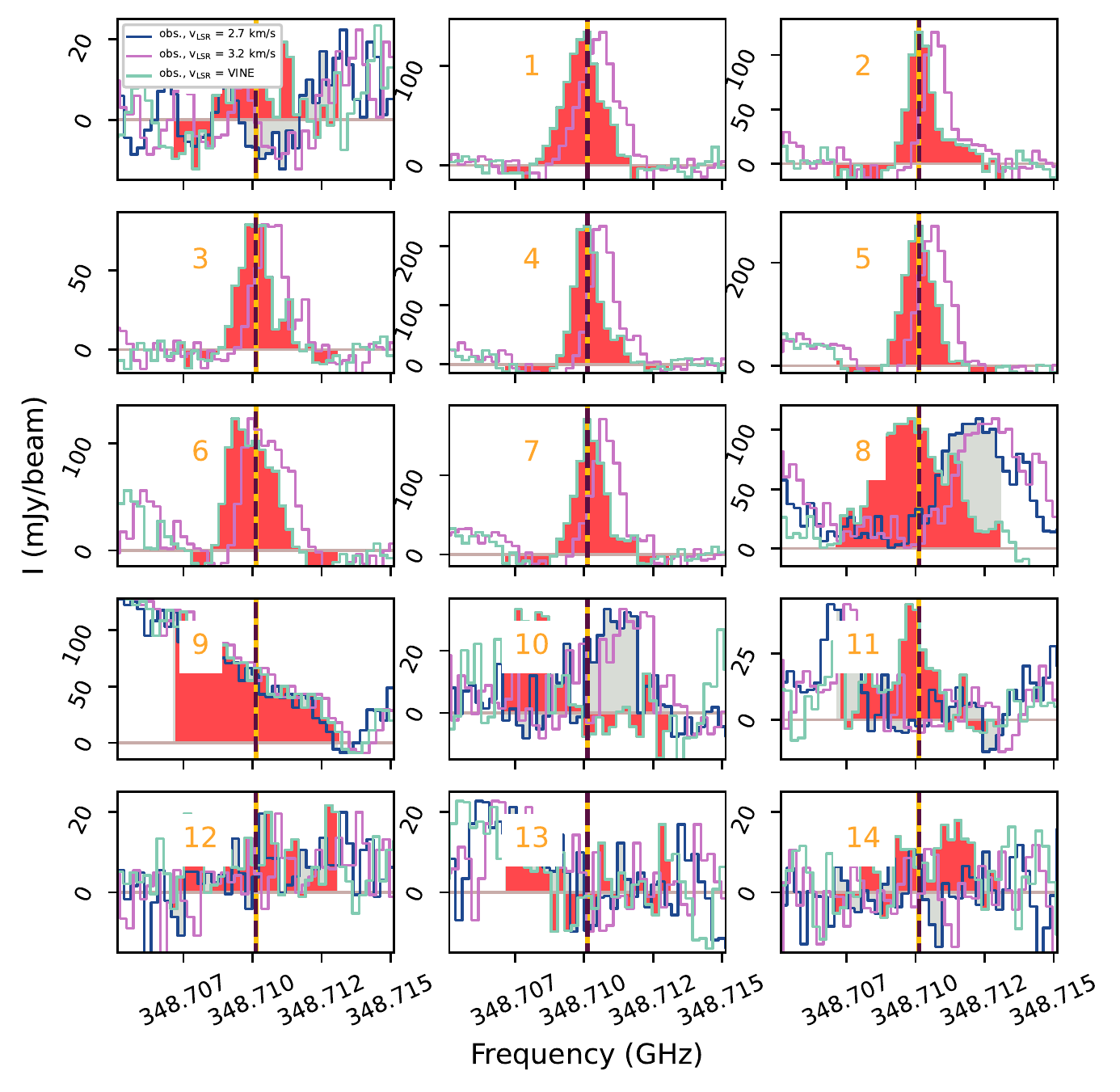}
	\caption{Spectra of the $348.710$~GHz line at 15 different position that are numbered and displayed in Fig.~\ref{fig:CD2HOH_348.7101234_mom0_2.7_multi}. The different colored curves correspond to different $v_{\text{LSR}}$ being applied. Blue corresponds to $v_{\text{LSR}}=2.7$~km~s$^{-1}$ that is most appropriate for source B, purple corresponds to $v_{\text{LSR}}=3.2$~km~s$^{-1}$ that is most appropriate for source A, and turquoise corresponds to the position-dependent  $v_{\text{LSR}}$ of the VINE method. The red region shows the region being integrated for the VINE map in Fig.~\ref{fig:CD2HOH_348.7101234_vine_multi_paper}. The gray region shows the region being integrated during the classical (non-VINE) method in Fig.~\ref{fig:CD2HOH_348.7101234_mom0_2.7_multi} (near source B it is fully overlapping with the red region).}
	\label{fig:CD2HOH_348.7101234_multi}
\end{figure*}

\begin{figure*}
	\centering
	\includegraphics[height=0.45\vsize]{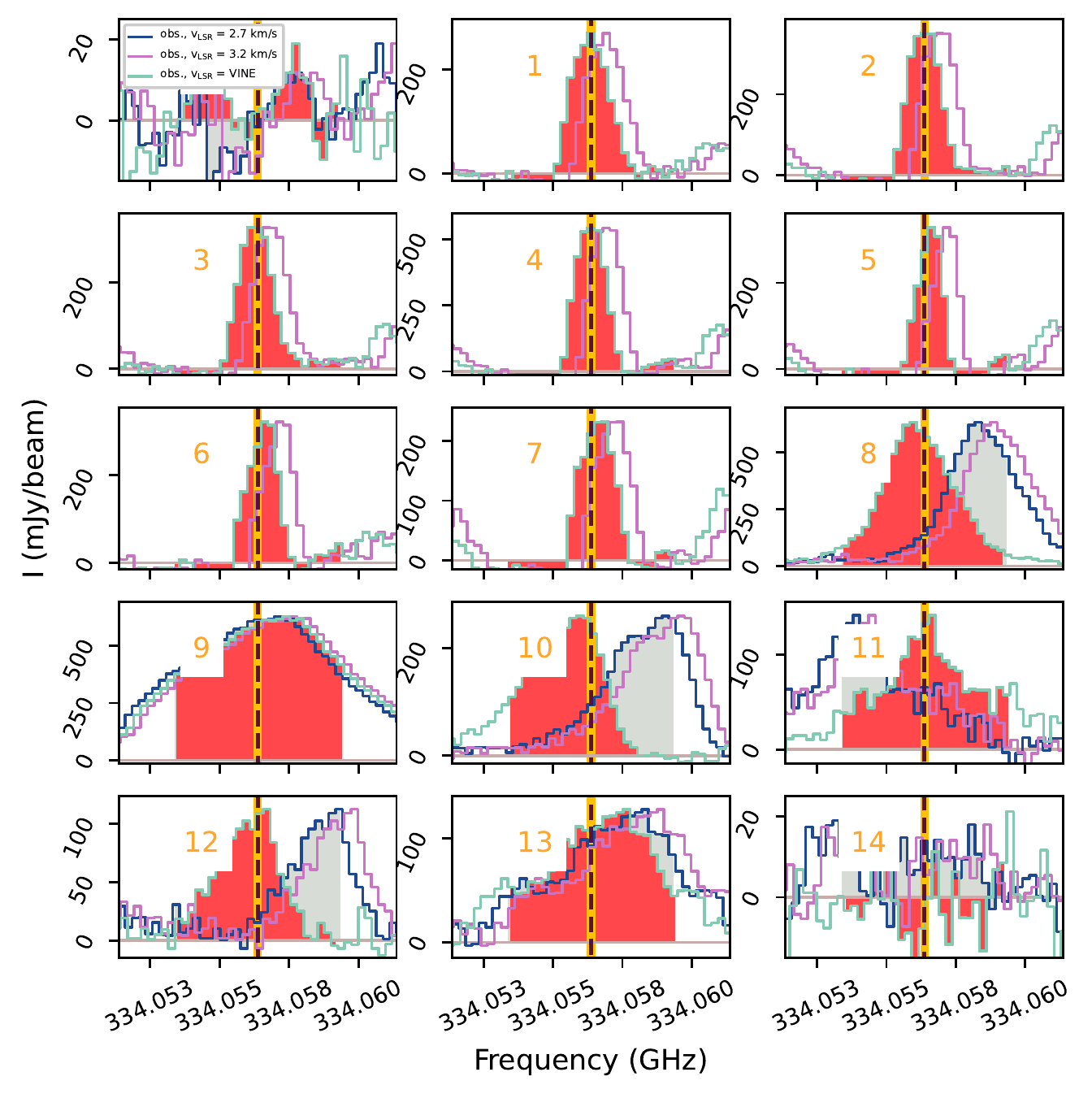}
	\caption{Spectra of the $334.056$~GHz line at 15 different position that are numbered and displayed in Fig.~\ref{fig:CD2HOH_334.056369_mom0_2.7_multi}. The different colored curves correspond to different $v_{\text{LSR}}$ being applied. The different colored curves correspond to different $v_{\text{LSR}}$ being applied. Blue corresponds to $v_{\text{LSR}}=2.7$~km~s$^{-1}$ that is most appropriate for source B, purple corresponds to $v_{\text{LSR}}=3.2$~km~s$^{-1}$ that is most appropriate for source A, and turquoise corresponds to the position-dependent  $v_{\text{LSR}}$ of the VINE method. The red region shows the region being integrated for the VINE map in Fig.~\ref{fig:CD2HOH_334.056369_vine_multi_paper}. The gray region shows the region being integrated during the classical (non-VINE) method in Fig.~\ref{fig:CD2HOH_334.056369_mom0_2.7_multi} (near source B it is fully overlapping with the red region).}
	\label{fig:CD2HOH_334.056369_multi}
\end{figure*}

Figs.~\ref{fig:CD2HOH_348.7101234_mom0_2.7_multi} and~\ref{fig:CD2HOH_334.056369_mom0_2.7_multi} have $14$ individual positions labeled. The corresponding line profiles at these positions are shown in Figs.~\ref{fig:CD2HOH_348.7101234_multi} and~\ref{fig:CD2HOH_334.056369_multi}, respectively. A closer inspection of position $\#9$, which corresponds to on-source A, shows that there is blending with a neighboring line in the case of the $348.710$~GHz line (which is only weakly emitting near source B) and that the $25$ spectral bins used for integration still do not fully grab all the emission in the $334.056$~GHz line (but a larger range would start including emission from nearby lines seen in the spectrum of source B). This illustrates that moment 0 maps, even VINE maps, have limitations when it comes to accurately showing the emission near source A. Potentially only much higher spatial resolution observations can disentangle the different components of source A and allow more narrow lines to be observed that would be suitable for the creation of moment 0 maps.

\clearpage

\section{Statistical corrections to abundance ratios for D/H ratio derivations}
\label{statistical_corrections}
The probability of a deuterium atom replacing a hydrogen atom at a specific location in a certain functional group ($-$XH$_{n}$) is a statistically independent event. As presented in appendix~B of \citet{Manigand2019}, the number of indistinguishable combinations with $i$ deuterium atoms at $n$ potential sites in a specific functional group is the statistical correction that must be applied to the abundance ratio of two isotopologs to obtain the molecular deuteration fraction (called the D/H ratio of a molecule) in a specific functional group. The D/H ratio depends on whether it is being considered relative to an isotopolog without deuterium in that functional group or relative to one that already contains $j$ deuterium atoms in that functional group as follows:
\begin{equation}
\frac{\text{XH}_{n-i}\text{D}_{i}}{\text{XH}_{n}} = \binom{n}{i} \left( \frac{\text{D}}{\text{H}}\right)_{\text{XH}_{n}}^{i},
\end{equation}
\begin{equation}
\frac{\text{XH}_{n-i}\text{D}_{i}}{\text{XH}_{n-j}\text{D}_{j}} = \frac{\binom{n}{i}}{\binom{n}{j}} \left( \frac{\text{D}}{\text{H}}\right)_{\text{XH}_{n-j}\text{D}_{j}}^{i-j}, i>j\geq0,
\end{equation}
where $\binom{n}{i} = \frac{n!}{i!(n-i)!}$. For example:
\begin{align}
\frac{\text{CH}_{2}\text{DOH}}{\text{CH}_{3}\text{OH}} &= 3 \left( \frac{\text{D}}{\text{H}} \right) _{\text{CH}_{3}\text{OH}},\\
\frac{\text{CHD}_{2}\text{OH}}{\text{CH}_{3}\text{OH}} &= 3 \left( \frac{\text{D}}{\text{H}} \right) _{\text{CH}_{3}\text{OH}}^{2},\\
\frac{\text{CH}_{3}\text{OD}}{\text{CH}_{3}\text{OH}} &= \left( \frac{\text{D}}{\text{H}} \right) _{\text{CH}_{3}\text{OH}}.
\end{align}

\section{Deuterated species detected on large-scales around IRAS~16293-2422}
\label{excluded_deuterated_species}
Deuterated species that are thought to trace the large-scale circumbinary envelope or even the larger surrounding cloud have been omitted from Fig.~\ref{fig:DH_plot_protostars}. Specifically, these are: DNO (tentatively claimed in absorption towards source A, \citealt{Chandler2005}; but noting that HNO has not been detected at the one-beam offset position from B, \citealt{Coutens2019b}), DCN \citep{Lis2002, vanDishoeck1995, Kuan2004, Takakuwa2007, Caux2011, Jorgensen2011, Wampfler2014, vanderWiel2019}, DNC \citep{vanDishoeck1995, Caux2011}, DC$_{3}$N (detected as part of TIMASSS without a derived column density, \citealt{JaberAl-Edhari2017}, but not detected at the one-beam offset position from B, \citealt{Calcutt2018a}), DC$^{15}$N \citep{Drozdovskaya2019}, ND \citep{Bacmann2010}, N$_{2}$D$^{+}$ \citep{Castets2001, Lis2002, Lis2016, Murillo2018b}, H$_{2}$D$^{+}$ \citep{Stark2004, Bruenken2014}, D$_{2}$H$^{+}$ \citep{Harju2017}, C$_{2}$D \citep{vanDishoeck1995, Caux2011}, OD \citep{Parise2012}, DCO$^{+}$ \citep{WoottenLoren1987, vanDishoeck1995, Lis2002, Stark2004, Caux2011, Jorgensen2011, Lindberg2017, Quenard2018a, Murillo2018b}, D$^{13}$CO$^{+}$ \citep{Caux2011, Quenard2018a}, c-C$_{3}$HD \citep{vanDishoeck1995, Caux2011, Majumdar2017}, c-C$_{3}$D$_{2}$ (labeled in \citealt{MartinDomenech2016a}), DCS$^{+}$ (tentatively detected towards source A, \citealt{Jorgensen2011}, but not detected at the one-beam offset position from B, \citealt{Drozdovskaya2018}). Finally, CH$_{2}$CDCN is omitted, as it was only tentatively detected by \citet{Huang2005} and was searched for, but not detected in the ALMA-PILS data (H. Calcutt priv. comm.).

\end{appendix}


\end{document}